\documentclass[1p]{elsarticle}

\usepackage{lineno,hyperref}
\modulolinenumbers[5]
\usepackage[dvipsnames,table,xcdraw]{xcolor}
\usepackage{enumitem}
\usepackage{empheq}

\usepackage{tikz}

\usepackage{booktabs}

\usepackage{array} 

\usepackage{pdflscape}

\usepackage{natbib}
\usepackage{graphicx}
\usepackage{multirow}
\usepackage{relsize}
\usepackage{derivative}
\usepackage{booktabs}

\usepackage[utf8]{inputenc}
\usepackage[normalem]{ulem}
\usepackage{amsmath}

\journal{Structural and Multidisciplinary Optimization}
\usepackage{amsmath,amsfonts,amssymb,amsthm,amsbsy,amsmath,bm}

\begin{document}

\begin{frontmatter}

\title{A practical review on promoting connectivity in topology optimization} 

\author[KUL,FM]{Vanessa Cool\corref{mycorrespondingauthor}}
\ead{vanessa.cool@kuleuven.be}

\author[DTU,CAMM]{Niels Aage}

\author[DTU]{Ole Sigmund}

\address[KUL]{KU Leuven, Department of Mechanical Engineering, Division LMSD, Celestijnenlaan 300 - box 2420, Heverlee, Belgium}
\address[FM]{Flanders Make@KU Leuven, Belgium}
\address[DTU]{Technical University of Denmark, Department of Civil and Mechanical Engineering, Nils Koppels Allé, Building 404, 2800 Kongens Lyngby, Denmark}
\address[CAMM]{Technical University of Denmark, Centre for Acoustic-Mechanical Micro Systems, 2800 Kongens Lyngby, Denmark}

\begin{abstract}
Topology optimization facilitates the automated design of high-performance structures across various engineering fields but, if unconstrained, often produces designs that are complex and difficult to manufacture.\
A key attribute of the resulting designs is \textit{connectivity}, which involves controlling the presence of solid and/or void islands of material.\
This manuscript provides a comprehensive overview of existing connectivity constraints developed for continuous design representations and highlights their advantages and limitations in influencing design outcomes and performance.\
The review further includes a practical comparison of five different connectivity constraints using a topology optimization framework for sandwich panels that balances acoustic and structural performance.\
With Pareto-front analyses, the constraints are evaluated based on computational cost, monotonicity, parameter dependency, and their impact on the optimized designs, their performance, and underlying dynamics.\
From the comparison, practical insights and rule of thumbs have been derived.\
The findings emphasize the critical role of selecting appropriate connectivity constraints, given their significant effect on the optimization results.\
\end{abstract}


\end{frontmatter}


\section{Introduction}
Topology optimization is nowadays recognized as a powerful tool to design high-performant structures in various fields of engineering~\cite{bendsoe1995optimization}.\
It is a numerical design approach which enables the creation of optimized designs while maximizing or minimizing certain performance metrics, without specifying an initial material layout.\
Over the years, several branches have emerged within topology optimization, largely divided into methods which use discrete or continuous underlying design variables to represent the topology~\cite{sigmund2013topology}.\
Although designs obtained using an unconstrained topology optimization framework achieve superior performance, the topologies are generally complex, necessitating extra control measures on the material layout, e.g.\ length-scale, separation of members, presence of holes, connectivity etc., already during the optimization.\

\begin{figure*}
	\centering
	\includegraphics[width = \linewidth]{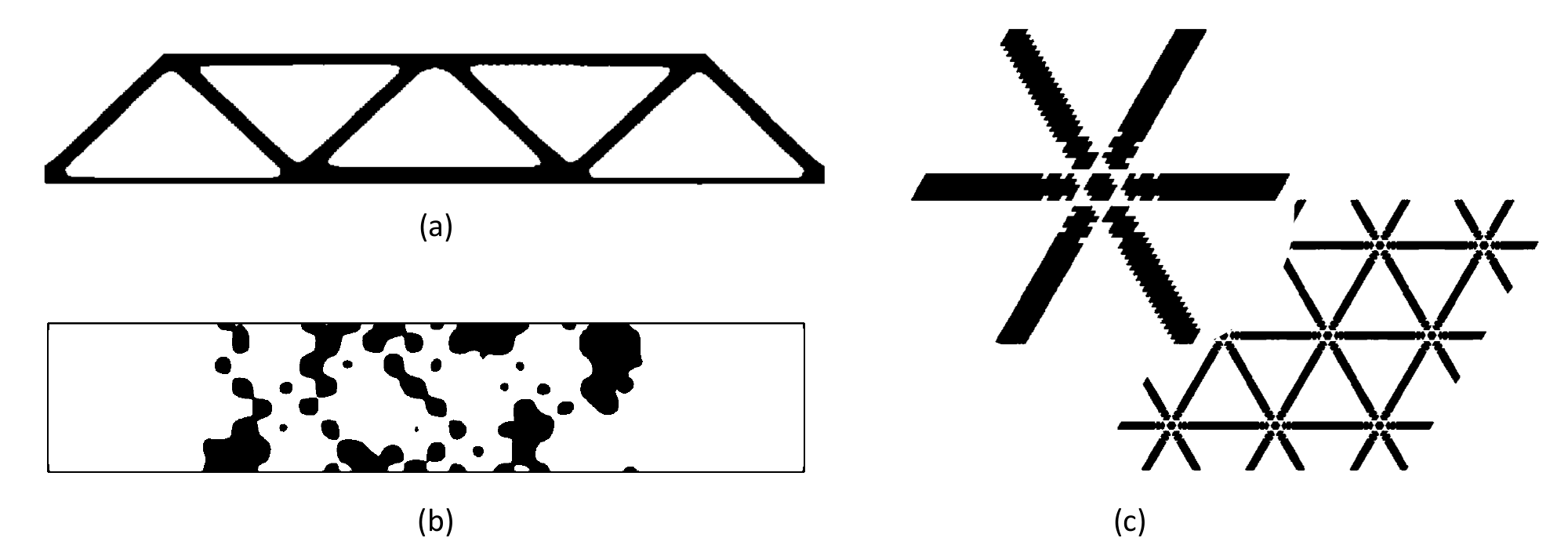}
	\caption{Examples showing the need to control the connectivity in a topology optimization context.\ a)~Structural compliance optimization of the reference MBB beam showing the typical internal voids.\ Figure adapted from~\cite{bendsoe1995optimization}.\ b)~Result of a vibro-acoustic optimization framework for acoustic broadband filters showing floating islands of material.\ Figure adapted from~\cite{dilgen2024topology}.\ c)~Result of a single-phase bandgap maximization optimization (unit cell and periodic structure) showing an island of material.\ Figure adapted from~\cite{halkjaer2006maximizing}.\ }
	\label{fig:lit}
\end{figure*}

One of the most important characteristics to control is the \textit{connectivity} of the layout during the optimization.\
The connectivity of a design is determined by the presence of disconnected members, void islands or solid islands of material.\
If no void or solid islands are present, one denotes the structure as \textit{simply-connected}.\
The presence of enclosed holes in the optimized structures was already evident from the first works in the field of topology optimization looking towards stiffness optimization, cf.~the typical MBB-beam example in Fig.~\ref{fig:lit}a.\
It is generally known that these holes are desirable from a stiffness perspective, but lead to more complicated designs in terms of manufacturing.\
In 3D, these enclosed voids can be impossible to manufacture, even with additive manufacturing~\cite{swartz2022manufacturing}.\
At the other side of the spectrum, issues with solid islands of material started to appear when other objectives than structural stiffness were of interest.\
Fig.~\ref{fig:lit}b shows an example of the optimization of an acoustic waveguide and Fig.~\ref{fig:lit}c shows an example of an elastic bandgap maximization in which both solid islands of material are natural to appear since they improve the performance.\ 
These free-floating structures are, however, non-feasible in reality.\
Including connectivity requirements is thus essential to obtain high-performant structures which are manufacturable as well.\
Moreover, when combined with a topology optimization framework, which uses continuous underlying design variables, creativity is necessary to determine the \textit{connectivity} during the optimization since no explicit knowledge on the material boundaries is present.\
Over the years, many different connectivity constraints have been presented in the literature which this review aims to coherently present and compare.\
The review focuses on connectivity control methods which are applicable for optimization methodologies in which the design is represented with underlying design variables which are continuous in nature, i.e.~not limited to either zero or one.\

\newpage
Although the effectiveness of these different constraints is widely discussed separately in the literature, a direct comparison between the different constraints and the impact the constraints can have on the resulting designs and their performances have not been presented before.\
This work, therefore, provides a practical comparison of five selected connectivity constraints by applying them to the same optimization problem.\
This enables a direct comparison.\
The design of a sandwich panel is chosen as example problem.\ 
Due to the nature of this problem, i.e.\ the presence of multiple physics and the inclusion of dynamics, the optimized designs are prone to be disconnected.\ Hence, this problem is well-suited for studying various connectivity constraints.\
Moreover, in the example, the connectivity characteristic is implicitly linked to the structural performance of the resulting design, i.e.\ better performance is obtained with disconnections.\
The five selected constraints are based on the static compliance, self-weight compliance, mechanical eigenvalues, virtual temperature method and spectral graph theory.\
An elaborate comparison and discussion is provided between the constraints on different aspects: (i)~the computational cost and tunability, i.e.~dependency of changes in the tolerance selection, of the constraints, (ii)~the consistency between the different constraints, (iii)~the impact the constraints have on the resulting designs and their performances, and (iv)~the monotonicity in the performance change when varying the constraint parameters.\

The rest of the manuscript is structured as follows.\ 
Section~\ref{sec:def} starts with defining \textit{connectivity} in a topology optimization context and briefly discusses peripheral research areas which are not elaborated in detail in this review.\ 
Afterwards, Sec.~\ref{sec:lit_cont} elaborates on the different connectivity constraints for continuous design variables.\ 
In Sec.~\ref{sec:comp}, a practical comparison of five different connectivity constraints is discussed with an application case.\
The review ends with the conclusions and lessons learned in Sec.~\ref{sec:concl}.\

\section{Definition of connectivity constraints in topology optimization}
\label{sec:def}
This section discusses the definition of \textit{connectivity} in topologies after which an overview of peripheral research areas is provided with corresponding sources.\

\subsection{Property of connectivity}
\label{subsec:prop_conn}
\begin{figure*}
	\centering
	\includegraphics[width = \linewidth]{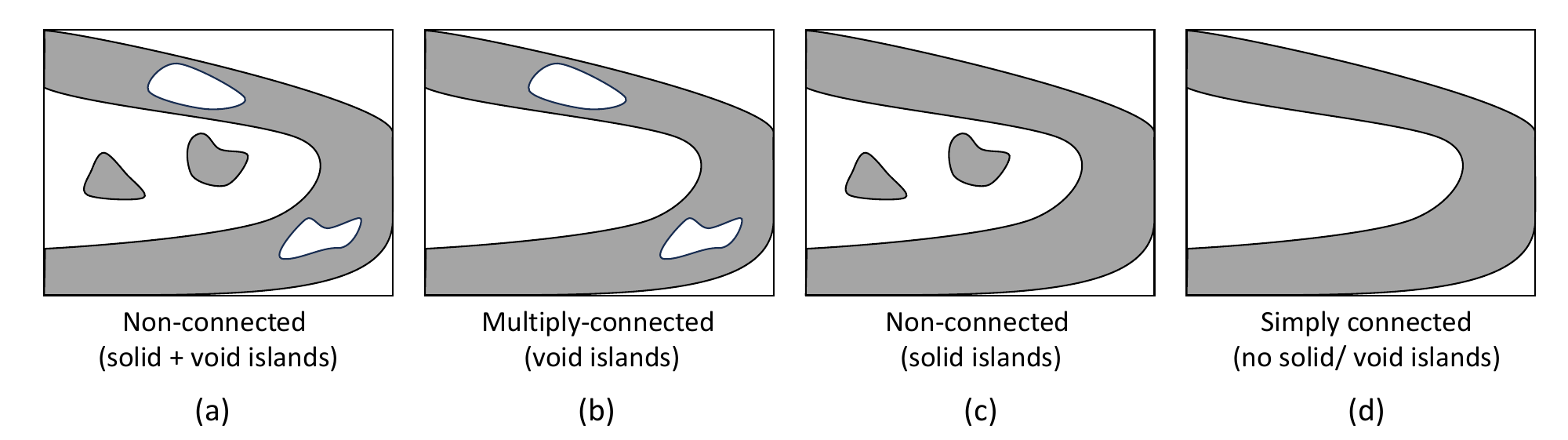}
	\caption{Definitions of connectivity in topology optimized designs.\ Example of a)~a non-connected design, b)~a multiply-connected design, c)~a non-connected design and d)~a simply connected design.}
	\label{fig:conn_def}
\end{figure*}

In order to define \textit{connectivity} in a material layout, it is easiest to look at a design which would be defined as \textit{non-connected}.\
An example is shown in Fig.~\ref{fig:conn_def}a.\
In the design, both solid islands and void islands can be seen.\
\textit{Solid islands} consist of different agglomerations of solid materials which are not connected through other solid material with the rest of the structure.\
In the literature, different names are used e.g.~\textit{islands of solid material}, \textit{isolated components}, \textit{free floating structures}, etc.\
On the other hand, void islands are regions of void space which are surrounded completely by solid material.\ 
These are also sometimes denoted as \textit{holes}, \textit{void islands}, \textit{enclosed void spaces} or \textit{enclosed voids}.\

Based on these two definitions, different types of connectivity constraints can be formulated.\
A first type of constraints tries to avoid the presence of solid islands, while void islands are allowed.\ 
An example of these type of structures is shown in 
Fig.~\ref{fig:conn_def}b and is also commonly denoted as \textit{multiply-connected} or \textit{fully-connected} topologies.\
A second type of constraints tries to avoid the presence of void islands, while not explicitly constraining the solid islands, cf.~Fig.~\ref{fig:conn_def}c.\
Note that in particular cases, e.g.~compliance optimization, this will results in a connected design since the solid islands do not have a practical purpose.\
Finally, both solid and void islands could be constrained leading to what is called \textit{simply connected} designs, as visualized in Fig.~\ref{fig:conn_def}d.\
This can be seen as a special case of the \textit{multiply-connected} or \textit{fully-connected} design.\

\begin{figure*}
	\centering
	\includegraphics[width = \linewidth]{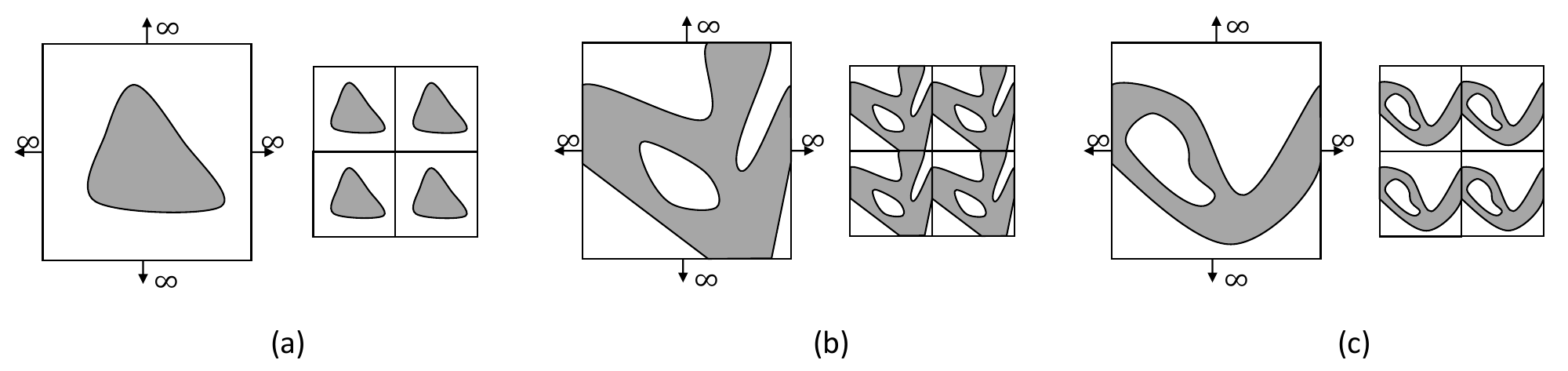}
	\caption{Note on the definitions of connectivity in periodic structure designs.\ Each time the reference domain is shown with a component consisting of two times two unit cells next to it.\ a)~Example of a design which is simply connected within the reference domain, but non-connected across unit cells. b)~Example of a design which is both multiply-connected on the reference domain and component level leading to a self-supporting structure.\ c)~Example of a design which is multiply-connected on the reference domain but not self-supporting on the component level since only horizontal connectivity is achieved. }
	\label{fig:conn_def_per}
\end{figure*}

A note should be made about the terminology of connectivity in periodic structures.\
Periodic structures are mostly defined by a unit cell (UC) or supercell which is repeated in some lattice directions.\
During the optimization, commonly only the reference unit cell or supercell is considered.\
Due to this definition, a difference exists in the connectivity on the reference domain and the full infinite periodic structure.\
Fig.~\ref{fig:conn_def_per} gives an example.\
Fig.~\ref{fig:conn_def_per}a displays a simply connected design on the reference domain (left), while this results in solid islands of material when considering multiple unit cells (right).\
Fig.~\ref{fig:conn_def_per}b displays a real fully connected design in the periodic sense, it is both connected on the unit cell level as the infinite periodic structure.\
In the literature, the term \textit{self-supporting} structure is used for the latter to differentiate both types of topologies~\cite{swartz2022manufacturing}.\
Note that obtaining a \textit{self-supporting} structure entails that the structure should be connected in both the horizontal and vertical direction.\ 
An example of a case violating the vertical connectivity is shown in Fig.~\ref{fig:conn_def_per}c.\

\subsection{Peripheral research related to the review}
\label{subsec:per_research}
This review focuses solely on connectivity constraints in topology optimization which deals with continuous design variables while \textit{connectivity} is defined as above, meaning either eliminating solid and/or void islands or achieving simply connected designs.\
There are, however, many other research directions within the scope of topology optimization which are closely related to the notion of connectivity.\
To maintain the focus of the review, only a brief introduction of these subgroups of research is provided here to give the reader a starting point for further literature studies:
\begin{itemize} [leftmargin=*]

    \item[-] \textit{Manufacturing constraints} - 
    The need for connectivity in the optimized topology has its nature in the manufacturablity of the design.\ Including manufacturing constraints for specific processes, such as additive manufacturing, casting, molding etc. has been a field of interest ever since topology optimization made its impact on industry in the early 2000s.\ The reader is referred to the available review papers in the literature for an overview regarding manufacturing constraint in topology optimization, e.g.~\cite{sutradhar2017incorporating,jihong2021review,liu2016survey,bayat2023holistic,vatanabe2016topology} and references therein.\
    Obtaining connectivity by avoiding enclosed voids or solid islands of materials is embedded indirectly in other manufacturing constraints, e.g.~ensuring certain parting directions for casting means no solid islands of material can occur.\ The specific manufacturing constraints are however more conservative than the connectivity constraints discussed in this review.
    
    \item[-] \textit{Regularization methods} - 
    In density-based topology optimization, regularization is needed to ensure the design problem is mesh independent.\ A special case of this concerns the so-called checkerboards.\ Checkerboarding is a known problem in which alternating solid and void elements occur.\ Several methods have been proposed to solve this issue,~e.g.~constraints on the perimeter or design variable gradient~\cite{haber1996new,niordson1983optimal}, regularization filters~\cite{bourdin2001filters} etc.\ The checkerboarding could be seen as a special case of issues with the connectivity of the design and solutions for the issue are in some cases even called \textit{connectivity search schemes}~\cite{wang2006enhanced}.\ Since the checkerboarding is mainly due to numerical problems, the solutions are not seen as connectivity constraints in this work and are not considered further in this overview.\ 
    Similarly, spurious small isolated islands of solid material can occur in structural optimization frameworks using the XFEM, cutFEM or level set methods.\ Solutions for this problem exist, e.g.~\cite{wei2010study,makhija2014numerical,villanueva2014density,villanueva2017cutfem}, and are not further considered as connectivity constraints in this review.\
    
    \item[-] \textit{Path searching} - 
    Many of the below discussed connectivity constraints which focus on eliminating enclosed voids have as a motivation the powder removal when making the component with additive manufacturing.\ 
    There are however two families of methods that achieve this, connectivity constraints which are considered during the optimization and \textit{path searching methods} which are only used in post-processing to design channels between the enclosed voids of an already obtained design.\ The latter is not considered further in this overview since the connectivity is not embedded directly in the optimization.\ Interested readers are referred to the literature for details on the post-processing methods, e.g.~\cite{yang2023differentiable,wei2019channel,wang2022topology} and references therein.\
    
    \item[-] \textit{Connectivity in multi-scale topology optimization} - 
    Multi-scale topology optimization is the field of structural optimization which optimizes the structure across different length scales.\ 
    Both the component on the macrolevel and the infill or microstructure is optimized at once.\
    The reader is referred to the review paper of Wu et al.~\cite{wu2021topology} for an overview.\
    In multi-scale topology optimization, it is a well-known problem that care should be taken to avoid a poorly connected microstructure.\
    Several methods have been proposed to avoid this issue, from physics-based constraints~\cite{zhou2008design,garner2019compatibility} to physics-independent connectivity constraints~\cite{du2018connecting,cramer2016microstructure}.\
    The field of de-homogenization achieves well-connected mono-scale designs by conformal mapping techniques, see e.g.~\cite{pantz2008post,woldseth2024phasor}.\
    The special measures to assure connectivity in multi-scale approaches are not further detailed in this review.\

    \item[-] \textit{Discrete vs. continuous design variable field} - 
    Both connectivity constraints which require discrete design variables and which work for continuous design variables have been presented.\
    The discrete design variable methods require the underlying design to be represented with a zero-one design variable field.\ 
    Due to the black and white nature of the underlying topology, an explicit definition of the connectivity can be identified.\
    For example, applying a Bi-directional Evolutionary Structural Optimization (BESO) optimization framework, Xiong et al.~\cite{xiong2020new} presented a method to avoid enclosed void islands using a shortest path algorithm.\ This was adapted using a genetic algorithm by Liu et al.~\cite{liu2022topology}.\
    These methods are also sometimes called structural connectivity control techniques.\
    The discrete methods are in contrast to methods which work for topologies represented with continuous design variables.\ 
    There, commonly threshold values in the connectivity constraint have to be used.\
    Since methods which work for a continuous design variable field are less restrictive than the ones which are presented assuming a discrete design variable field, only the former will be further detailed in this review.

    \item[-] \textit{Structural Complexity Control} (SCC) - A last field of research which can be connected to the connectivity control is denoted the \textit{Structural Complexity Control}/ SCC.\ 
    It takes the connectivity control a step further and tries to control the topology of the resulting design in terms of the geometric features or explicit control on the number of holes and shape of the holes.\
    The research field is often divided into implicit indirect methods and explicit direct methods.\ 
    The former family consists of filtering techniques, such as density filtering~\cite{bourdin2001filters}, and length-scale control methods, such as the robust method~\cite{wang2011projection} or geometrical constraint method~\cite{zhou2015minimum}.\ 
    The direct methods explicitly control the resulting topology.\
    Several methodologies have been presented for various optimization framework, e.g.~in the (B)ESO framework~\cite{kim2000method,zhao2020direct,han2021topological,he2022thinning,he2023hole}, using a moving morphable component (MMC) approach~\cite{zhang2017structural}, in the level-set method~\cite{zhang2017explicit,zhou2023hole}, using a density-based optimization framework~\cite{wang2022compl} etc.\
    Also works have been presented using topological invariants of the structure (i.e.~the Euler and Betti numbers), to explicitly control the topology, both for discrete variable topology optimization~\cite{liang2022explicit} and density-based topology optimization~\cite{zuo2023explicit}.\

\end{itemize}

\section{Connectivity constraints with continuous design variables}
\label{sec:lit_cont}
This section gives on an elaborate overview of the different connectivity constraints.\ 
After a general overview of the different constraints, firstly physics-based connectivity constraints are discussed, after which the section elaborates on geometry-based connectivity constraints.\
A final subsection provides an overview of the capabilities of all discussed constraints.\

\subsection{General overview}
\label{subsec:gen_overview}
\begin{figure*}
	\centering
	\includegraphics[width = \linewidth]{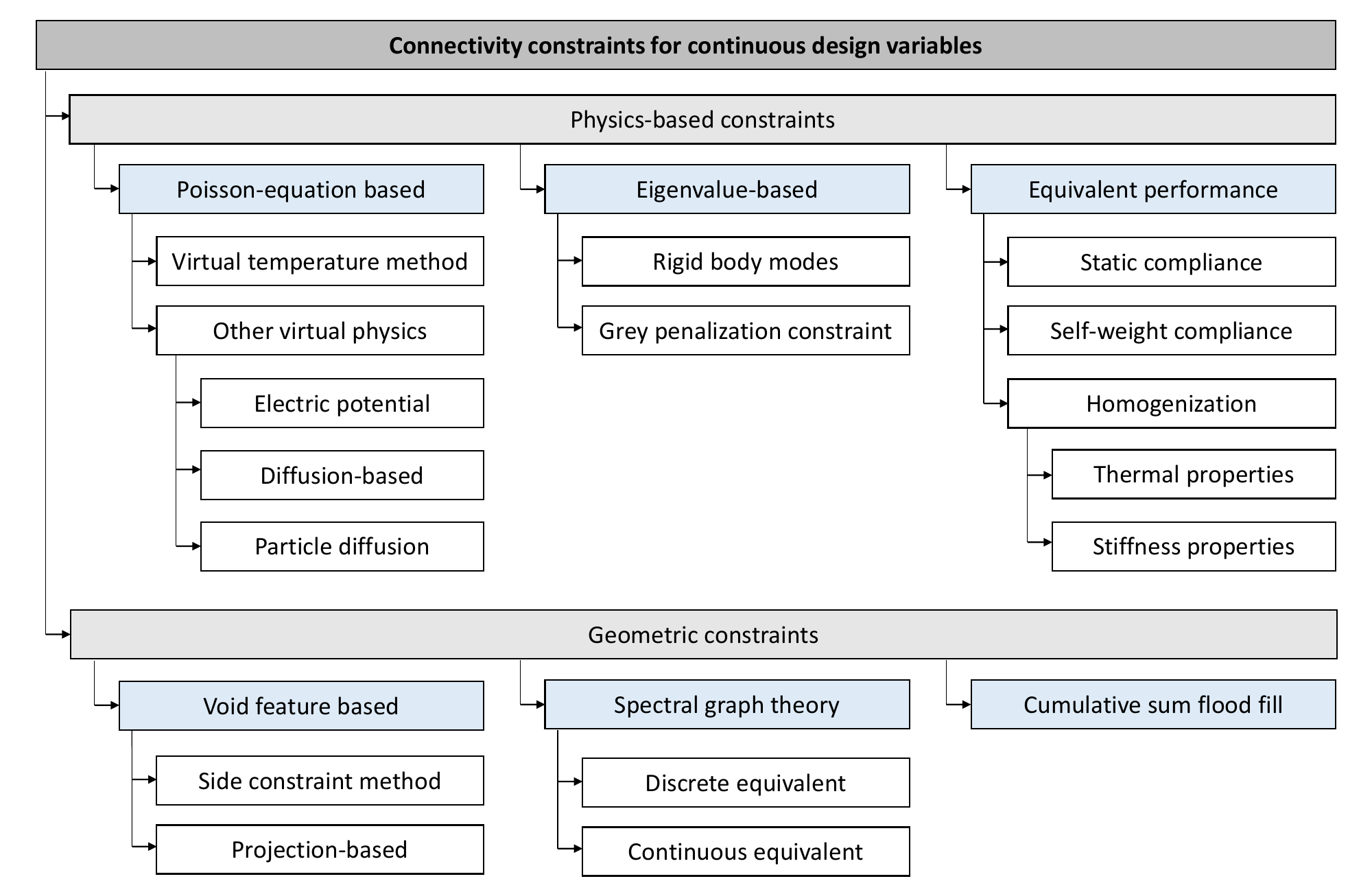}
	\caption{General overview of the different connectivity constraint which work for a continuous underlying design field and which are discussed in more detail in this review.\  }
	\label{fig:scheme}
\end{figure*}

This section gives a general overview on how the different connectivity constraints discussed in this review are divided into categories.\
This division is visualized in Fig.~\ref{fig:scheme}.\
In the family of connectivity constraints for continuous design variables, a division is made between physics-based and geometric constraints.\
Physics-based approaches use some kind of underlying physical phenomena to define the \textit{connectivity} of the design e.g.~conductivity or stiffness properties.\ 
They mostly require the solution of an extra PDE and give a more implicit control on the connectivity of the topology requiring tuned user-inputs.\
Geometric constraints on the other hand, use the explicit geometry of the design without the need of physical equivalences.\ 
They do not always require the computation of an extra PDE and can in some cases enable a more explicit control on the topology, c.f.~in terms of the number of void islands.\
These constraints are now further detailed in the following subsections.\

\subsection{Physics-based connectivity constraints}
\label{subsec:physics_constraint}
A first category of connectivity constraints, which can deal with continuous design variables, is denoted as physics-based connectivity constraints.\
As discussed before, these constraints control the connectivity by solving an extra problem and exploiting physical phenomena such as conductivity or stiffness properties.\
The category is divided into three subcategories which are further discussed below: (i)~Poisson-equation (diffusion) based scalar field constraints, (ii)~eigenvalue-based connectivity constraints and (iii)~equivalent-performance based constraints.\

\subsubsection{Poisson-equation based scalar field constraints}
\label{subsubsec:poisson_constraint}
A first set of physics-based connectivity constraints uses a virtual field which can be represented by the Poisson-equation to detect anomalies in the connectivity.\
The first works on promoting connectivity in topology optimization were based on this concept with the virtual field being a temperature field.\

\begin{figure*}
	\centering
	\includegraphics[width = \linewidth]{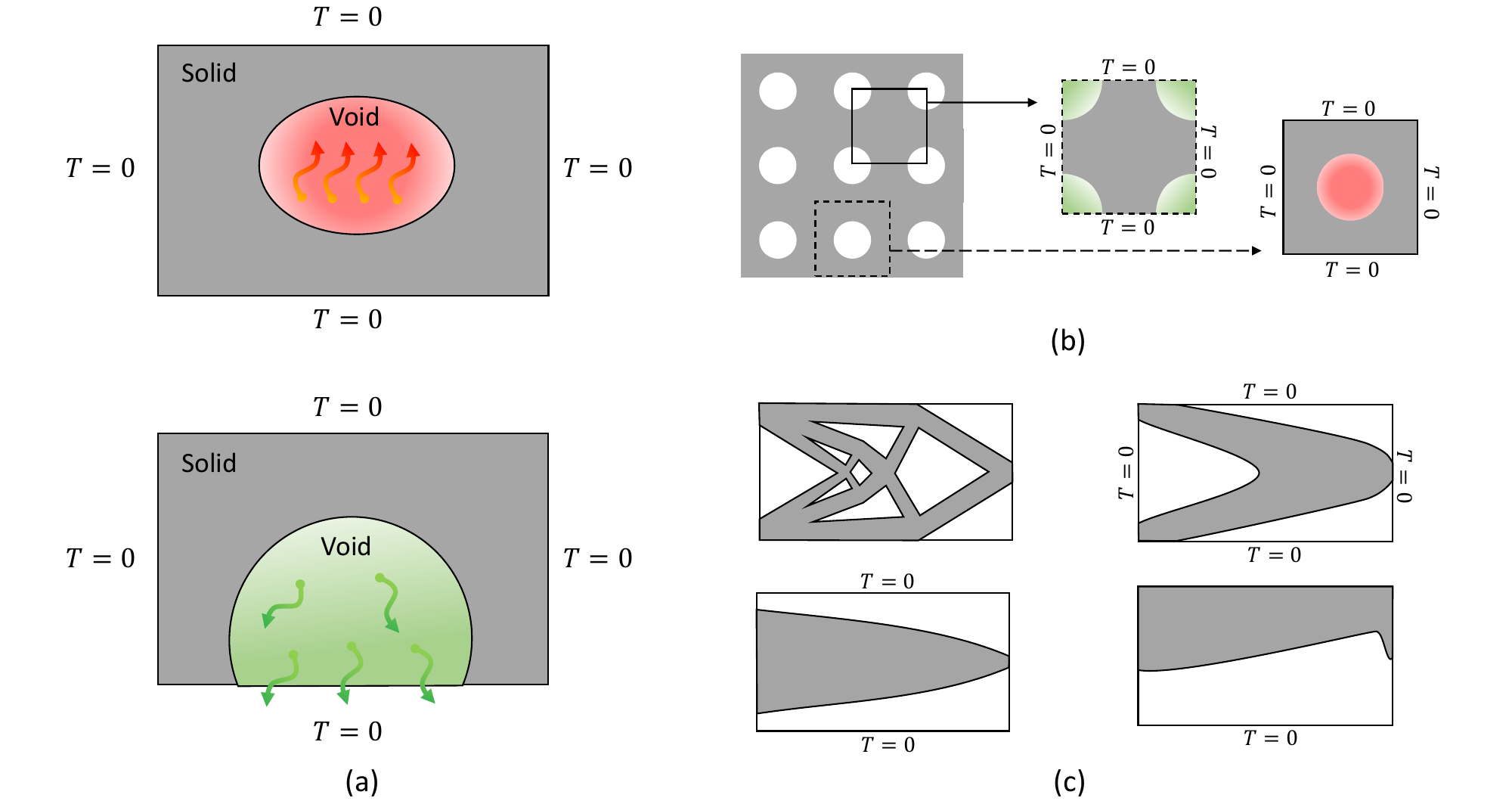}
	\caption{a)~Visualization of the concept of the virtual temperature method. At the top, a design with trapped heat is shown while at the bottom the heat of the void phase can escape through the boundaries.\ Figure adapted from~\cite{li2016structural}. b)~Visualization of why 2$^d$ unit cells should be considered when using the VTM for periodic structures.\ Figure adapted from~\cite{swartz2022manufacturing}.\ c)~Visualization to see the influence of the design by changing the Dirichlet boundary conditions.\ Figure adapted from~\cite{wang2020numerical}. }
	\label{fig:vtm}
\end{figure*}

\paragraph{Virtual temperature method}
The use of conductivity based constraints to ensure connectivity has been used for several decades, e.g.~\cite{sigmund1999optimality}.\ 
However, the  method was first formalized later with the introduction of the virtual temperature method.\
The virtual temperature method (VTM) was proposed simultaneously by different authors~\cite{liu2015identification,li2016structural,osanov2016topology} in order to avoid void islands or obtain simply-connected structures.\
The concept of the connectivity constraint is shown in Fig.~\ref{fig:vtm}a.\
A virtual temperature field is introduced to detect void islands.\
More specifically, a heat source and highly conductive material is added to the void elements while insulating material is added to the solid material.\
When solving the conductivity problem with a Dirichlet boundary condition on the temperature ($\mathrm{T}=0$) at the boundaries, the heat will increase in all void regions which are not connected to the boundary since the heat cannot escape.\
Based on the maximum obtained temperature ($\mathrm{T_{max}}$), isolated islands of void become detectable.\
The underlying conduction problem is governed by the Poisson-equation:
\begin{equation}
    \nabla(k \nabla \mathrm{T}) + \mathrm{Q} = \mathrm{0}, \hspace{1cm} \mathrm{T}=0 \hspace{0.3cm} \mathrm{on} \hspace{0.3cm} \Gamma
\end{equation}
in which $\mathrm{T}$ represents the temperature field, $k$ is the heat conductivity coefficient, $\mathrm{Q}$ is the added heat source and $\Gamma$ represents the boundary of the domain.\
The connectivity constraint then reads
\begin{equation}
   \mathrm{T_{max}} \leq \bar{\mathrm{T}}
\end{equation}
with $\bar{\mathrm{T}}$ being the threshold value for the allowed temperature.\
The difference between avoiding void islands or obtaining a simply-connected structure is in the choice of boundary conditions.\

The VTM is one of the most applied connectivity constraints, ranging from applications in the design of phononic crystals~\cite{jia2024maximizing}, nanophotonic devices~\cite{kuster2025inverse} to the application of large-aperture space telescopes~\cite{hu2017design} and electrodes with electroactive polymers~\cite{wallin2024connectivity}.\
Although originally proposed for density-based topology optimization, Yamada and Noguchi extended the concept also towards the level-set method~\cite{yamada2022topology}.\

\paragraph{Extension of the virtual temperature method}
Starting from the initial concept introduced by~\cite{liu2015identification,li2016structural,osanov2016topology}, different extensions have been proposed in the literature.\

Although the VTM was originally proposed for the detection of void islands, the method can easily be converted to enable the detection of solid islands of material~\cite{swartz2022manufacturing,wu2024method}.\ 
In the case of solid islands detection, a heat source and highly conductive material is put on the solid elements, while the void material is insulating material.\

The VTM was originally proposed with a heat sink at all the boundaries of the domain.\ 
The selection of the boundary conditions can, however, lead to different needs~\cite{donoso2019topology,wang2020numerical}.\ Fig.~\ref{fig:vtm}c gives an example showing the impact of selecting different boundary conditions while solving the same optimization problem.\ 
The VTM can also be adapted to enable a more strict molding or casting constraints with specific boundary conditions~\cite{sato2017manufacturability,li2018topology}.\

Swartz et al.~\cite{swartz2022manufacturing} have extended the VTM towards the use in periodic media.\
Since the optimization is based on the smallest non-repetitive part, care should be taken while using the constraint.\
The authors show that a minimum array of $2^d$, with $d$ the dimension of periodicity, should be considered while enforcing the virtual temperature connectivity constraint to avoid issues with dependencies in the unit cell design.\
This is visually represented in Fig.~\ref{fig:vtm}b.\

The original VTM is known to be highly dependent on the user-defined inputs such as the conductivity and maximum strength of the heat source and on the investigated geometry.\
It was even observed that the linear heat source could lead to nonphysical behavior, i.e.~negative temperatures, in the intermediate design field elements.\
For this reason, several adaptations have been proposed.\
Swartz et al.~\cite{swartz2022manufacturing} proposed an adapted interpolation for the linear heat source.\ 
Using a modified SIMP interpolation~\cite{du2007topological}, the heat source in the high volume fractions is heavily penalized.\
Another technique to avoid the high sensitivities towards the user-inputs was proposed by Luo et al.~\cite{luo2020additive} with the so-called Non-linear~VTM (NVTM).\ 
In this method, the linear heat source is adapted towards a nonlinear heat source $Q$:
\begin{equation}
     \mathrm{Q}(\mathrm{T}) = \frac{\mathrm{q}}{1+e^{\alpha(\mathrm{T}-\mathrm{T}_m)}},
\end{equation}
in which $\alpha$ and $\mathrm{T}_m$ are two positive scalar numbers.\
The idea behind the non-linear heat source is to make the method less sensitive to the tuning parameters.\
With the above non-linear heat source, the temperature in the enclosed voids is bounded by the value $\mathrm{T}_m$, enabling a uniform temperature in the enclosed voids.\ With this, the connectivity constraints can be written more generally without the need of a user-defined threshold: $\mathrm{T_{max}}<\mathrm{T}_m/2$.\

The NVTM is further extended towards the bi-directional NVTM by Gu et al.~\cite{gu2024structural}.\ 
Their method focuses on identifying enclosed voids by only adding virtual heating material to regions below a certain temperature threshold while adding heat-absorbing material in regions exceeding this threshold.\
The addition of the heat-absorbing material enhances the precision to detect enclosed voids.\
Luo et al.~\cite{luo2022topology} presented a method to directly identify regions which need infill-support material for the design of structures which could be made with additive manufacturing.\
They presented a filter based on the NVTM to separate enclosed voids and open regions.\
Also Huang et al.~\cite{huang2022thermal} used the NVTM to distinguish open and closed voids during the optimization of structures with design-dependent loads in a thermal-solid-fluid method.\

\paragraph{Other physics in the virtual domain}
Although the VTM is based on a temperature equivalence, the general principle could be extended to other equivalences as well.\
Donoso and Guest~\cite{donoso2019topology} presented the so-called \textit{virtual electric potential field}.\ 
While ensuring connectivity in two-phases for the design of piezo modal transducers, they proposed to adapt the virtual temperature field to the counterparts of the electrical potential, i.e.~the heat source and thermal conductivity become current source and electrical conductivity.\
Although the same governing Poisson-equation describes the underlying problems, the electric potential equivalent seemed more natural in their application case.\
Wang et al.~\cite{wang2020numerical} explicitly described the generality of the Poisson-equation based scalar field connectivity constraint and showed the efficiency of the electrostatic equivalence.\ 
They extended the electrostatic virtual connectivity constraint also to the more limiting casting constraint~\cite{wang2020topology} and combined it with stress-constraints to simultaneously optimize for connected designs with a certain structural strength~\cite{wang2021structural}.\

In topology optimization methods for fluid-structure interaction problems, isolated volumes of fluid surrounded by solid could lead to singular analysis problem~\cite{jenkins2016immersed}.\
To remedy this issue, Jenkins and Maute~\cite{jenkins2016immersed} used a method inspired by the VTM to detect and remove these islands in an XFEM based fluid-structure interaction problem.\
The free-floating volumes are identified by solving an additional diffusion (Poisson) model in which both an ambient convective flux and a prescribed boundary flux is applied.\
Also Behrou et al.~\cite{behrou2019adaptive} applied this technique in a density-based topology optimization framework for incompressible flow problems.\

All the above Poisson-equation based connectivity constraints require the solution of an additional FEM operation.\ 
These are typically computationally cheap since the virtual field makes use of a scalar design variable field.\
Sabiston and Kim~\cite{sabiston2020void} proposed an alternative first-principle physics-based approach to avoid enclosed voids based on the particle diffusion theory which does not use an additional FEM solution.\
Instead, the concept uses a diffusion mechanism computation in which each element has a diffusion particle which can move until it hits the boundary of the domain.\
Based on the time it takes the virtual particles to reach the boundary, enclosed voids can be avoided.\

\newpage
\subsubsection{Eigenvalue-based constraints}
\label{subsubsec:eig_constraint}
Topology optimization with the objective of maximizing or tuning of  eigenfrequencies of structures started with the seminal work of Pedersen~\cite{pedersen2000maximization} and is still an active field of research, e.g.~\cite{li2021topology,wu2020substructuring,giannini2022topology}.\

Over the years, the equivalence between solid islands/connectivity constraints in designs and the nature of the eigenfrequencies have been investigated leading to connectivity constraints based on mechanical eigenvalues.\

\begin{figure*}
	\centering
	\includegraphics[width = \linewidth]{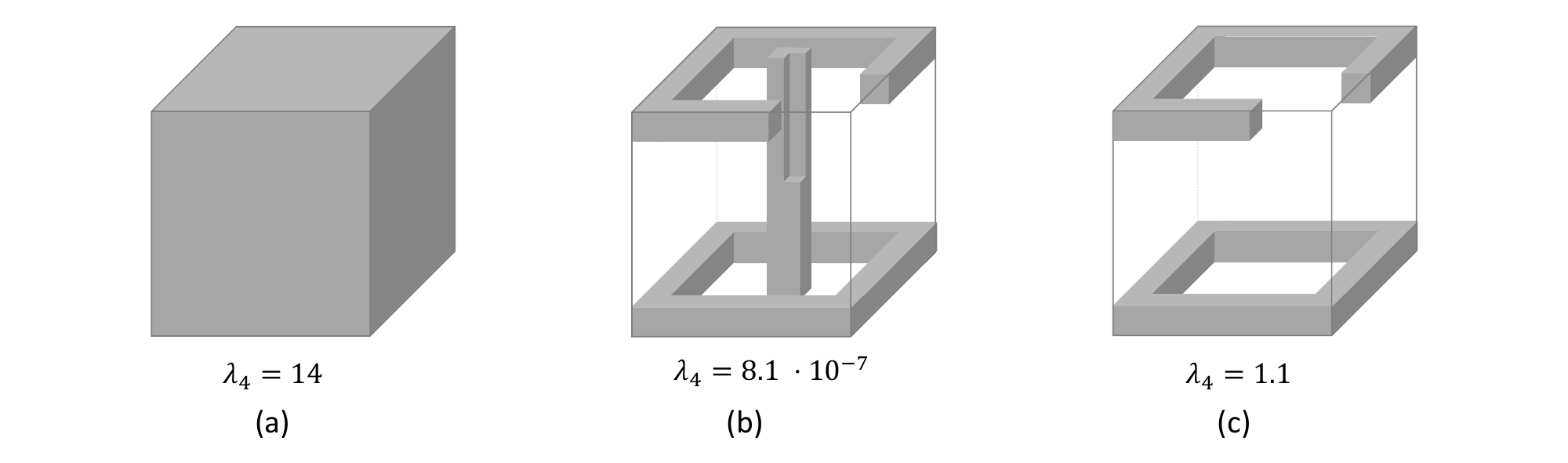}
	\caption{Example of why the eigenvalue constraint considering one UC and periodic boundary conditions can only assure connectivity within the UC and not across UCs.\ A 3D UC is considered with periodicity in the three direction.\ The fourth eigenvalue is applied for the connectivity constraint.\ The full solid case (a) has a non-zero fourth eigenvalue.\ The UC in (b) has a zero fourth eigenvalue since two structural bodies are detected when solving the eigenvalue problem with periodicity constraints.\ The UC in (c) however does not have a zero fourth eigenvalue since only one structural body is seen by the solved eigenvalue problem.\ So although the structure is connected on a UC level, the method does not detect the final structure is not self-supporting.\ Figure adapted and values taken from~\cite{swartz2022manufacturing}. }
	\label{fig:eig}
\end{figure*}

The work of Wang et al.~\cite{wang2011robust} is one of the first investigations exploiting the information of mechanical eigenfrequencies to avoid isolated islands of solid material in topology optimization.\
The study focused particularly on 2D photonic crystal waveguides using the robust formulation, but the proposed concept is generally applicable.\
The fundamental idea was that isolated components during the optimization would lead to zero-frequency rigid body modes of the investigated supercells.\
The authors, therefore, solved a solid mechanics eigenvalue problem with appropriate Dirichlet boundary constraints towards the eigenfrequencies $\lambda$:
\begin{equation}
    (\mathbf{K}_s-\lambda \mathbf{M}_s) \mathbf{q} = \mathbf{0} 
\end{equation}
with $\mathbf{K}_s, \mathbf{M}_s$ the FE structural stiffness and mass matrices and $\mathbf{q}$ the eigenmode.\ 
Now solid islands of materials can be eliminated by enforcing that eigenfrequency $j$ is larger than a certain lower bound $\bar{\lambda} > 0$:
\begin{equation}
    \lambda_j \geq \bar{\lambda}.
\end{equation}
Note that dependent on the applied boundary conditions, some rigid body modes may be physically meaningful, and hence, the lowest mode ($j$) to constrain can differ as a result of the chosen boundary conditions.\
This constraint was denoted the \textit{fundamental free vibration frequency constraint} by the authors.\
The effectiveness of the constraints was proven while it is noted that the algorithm is highly sensitive towards the choice of the parameter $\bar{\lambda}$.\
Swartz et al.~\cite{swartz2022manufacturing} recently extended the mechanical eigenvalue constraint towards the avoidance of solid islands in the design of 3D periodic structures.\
They applied periodic boundary conditions on the edges of the design without essential boundary conditions.\
This may ensure fully connected designs inside the unit cell while the self-supporting nature of the structures is not controlled.\ 
Fig.~\ref{fig:eig} gives an example taken from~\cite{swartz2022manufacturing} to exemplify this statement.\

Next to constraining the rigid body modes, another technique using eigenfrequencies to avoid weak connections was proposed by Giannini et al.~\cite{giannini2020topology} for the design of 2D in-plane MEMS gyroscopes.\
Weak gray connections are pushed towards void which causes disconnections and large changes in the eigenfrequencies.\
By introducing a new field of variables $\gamma_n$, determined with a Heaviside projection function to avoid intermediate gray values, the difference between the newly obtained eigenvalues $\omega_n$ and reference eigenvalues $\omega$ is bounded:
\begin{equation}
    1- \frac{\omega_n}{\alpha_g \omega} \leq 0,
\end{equation}
with $\alpha$ the lower bound parameter $0 \leq \alpha_g \leq 1$.\
The authors denoted these constraints as the \textit{gray penalization constraints}.\
Recently, this constraint has also been applied for multi-modal resonator designs~\cite{giannini2024topology}.\

\subsubsection{Equivalent performance-based constraints}
\label{subsubsec:eq_perf_constraint}
Next to the Poisson-equation and eigenvalue-based connectivity constraints, also equivalent performance-based constraints have been applied.\
Based on the equivalence between compliance performance of structures and the required connectivity which should ensure the desired stiffness by connecting loads to supports, several connectivity constraints have been presented.\

\paragraph{Static compliance}
\label{par:static}
From the start of the topology optimization research field, the static compliance has been a widely studied objective and is nowadays considered a solved problem.\
Over the years, the equivalence between obtaining a certain stiffness and the corresponding need to have connected designs without islands of material has been taken for granted as a straightforward way to achieve connectivity, when using a stiffness penalization scheme such as SIMP.\
From the onset of topology optimization, also compliant mechanisms were of interest.\ 
During the design of these structures, a certain static stiffness was sometimes desired between the considered input and output to avoid the typical one-node connected hinges~\cite{sigmund1997design,zhang2018topology,zhu2020design}.\
This can be seen as one of the initial uses of the static compliance to control the \textit{connectivity} of the design.\
Also in the work of Halkjaer et al.~\cite{halkjaer2006maximizing}, which is generally seen as one of the initial works in which solid islands of material were obtained, the authors indicated the idea of using the static stiffness of the design to obtain connectivity.\
The static compliance of a structure is obtained by solving the following solid mechanics problem:
\begin{equation}
\left\{
\begin{array}{ll}
\nabla \cdot \mathbb{C}(\xi) \left[ \nabla \mathbf{u} \right]  = \mathbf{F} \hspace{0.5cm}  \text{in}~\Omega \\ 
\mathbf{u} = 0 \hspace{2.4cm} \text{on}~\Gamma,
\end{array}
\right. 
\end{equation}
in which $\mathbb{C}$ is the elasticity tensor, $\mathbf{u}$ is the resulting displacement, $\xi$ is the design variable field and $\mathbf{F}$ is the external load while $\Gamma$ is the constrained boundary.\
From the resulting displacement and external applied force, the static compliance is obtained:
\begin{equation}
    \theta_{st} = \int_{\Omega} \mathbf{F} \cdot \mathbf{u}  d\Omega.
\end{equation}
The connectivity constraint then reads $\theta_{st} \leq \mu_{st}$ with $\mu_{st}$ the user-defined threshold.\
Note that care should be taken on how this is defined, since a fully connected design with a solid island of material can still have a high static stiffness depending on where the external force excites the structure.\
Hence, a volume penalization term minimizing the total used material is needed in order to avoid free floating material.\
Recent works which employ the static compliance constraint can still be found, e.g.~\cite{larsson2024topology,cool_TO_VA}.\

\paragraph{Self-weight compliance}
\label{par:self_weight}
Another performance-based constraint, which can control the connectivity of the design during the topology optimization, is the self-weight compliance.\
The investigation of structures under self-weight is an entire research field on its own with the fundamental work of Turteltaub and Washabaugh~\cite{turteltaub1999optimal}, Park et al.~\cite{park2003topology} and Bruyneel and Duysinx~\cite{bruyneel2005note}.\
As the static compliance constraint, the concept of the self-weight compliance constraint was not proposed by one manuscript, since it was applied over the years as a way to control the stiffness of the design.\
However, Swartz et al.~\cite{swartz2022manufacturing} recently explicitly proposed the self-weight compliance constraint to ensure self-supporting structures during the design of periodic media.\
The idea is that solid islands of material would experience large displacement under there self-weight and can therefore be detected by limiting the self-weight compliance.\
More specifically, similar to the static compliance, the following solid mechanics problem with now design-dependent body loads is solved:
\begin{equation}
\left\{
\begin{array}{ll}
\nabla \cdot \mathbb{C}(\xi) \left[ \nabla \mathbf{u}_i \right]  = -\xi \mathbf{e}_i \hspace{0.5cm}  \text{in}~\Omega \\ 
\mathbf{u}_i = 0 \hspace{2.9cm} \text{on}~\Gamma_i,
\end{array}
\right. 
\end{equation}
in which $\mathbf{e}_i$ is the body load in spatial dimension $i$ while $\Gamma_i$ is the corresponding boundary transverse to dimension $i$ which is constrained.\
From the different displacements $\mathbf{u}_i$, the self-weight compliance $\theta_{sw}$ is determined~\cite{swartz2022manufacturing}:
\begin{equation}
    \theta_{sw} = \sum_{i=1}^{N_d} \int_{\Omega} \xi \mathbf{e}_i \cdot \mathbf{u}_i  d\Omega,
\end{equation}
with $N_d$ the number of dimensions in which a force is applied.\
The connectivity constraint to avoid floating islands of material then reads $\theta_{sw} \leq \mu_{sw}$ with $\mu_{sw}$ the user-defined threshold.\
Note that, similar to the connectivity constraints for periodic media as discussed before, $2^d$ with $d$ the periodicity dimension, should be considered to avoid dependency on the unit cell choice.\
The self-weight compliance has shown its effectiveness to obtain self-supporting periodic media.\
It is, however, generally known that the self-weight compliance can cause issues with non-monotonous behavior and care should be taken with unwanted effects in the low density regions~\cite{bruyneel2005note}.\

\paragraph{Homogenization}
Numerical homogenization enables the calculation of the effective macroscopic properties of a periodic structures based on the smallest non-repetitive part, called the unit cell.\
An introduction into the field of homogenization can be found in several reference works~\cite{torquato2002random,bensoussan2011asymptotic,andreassen2014determine}.\
According to the classical homogenization theory, the equivalent or homogenized stiffness tensor $\mathbf{C}^h$ can be found by~\cite{allaire1997shape,andreassen2014determine}:
\begin{equation}
    \mathbf{C}_{ijkl}^h = \frac{1}{|\Omega|} \int_{\Omega} \mathbf{C}_{pqrs} (\epsilon_{pq}^{0(ij)}-\epsilon_{pq}^{(ij)})(\epsilon_{rs}^{0(kl)}-\epsilon_{rs}^{(kl)}) d\Omega, 
    \hspace{1cm}
    \epsilon_{pq}^{(ij)} = \frac{1}{2}(\chi_{p,q}^{ij} + \chi_{q,p}^{ij}),
\end{equation}
in which $|\Omega|$ is the volume, $\mathbf{C}_{pqrs}$ the local varying stiffness tensor, $\epsilon_{pq}^{0(ij)}$ is the pre-loading strain field and  $\epsilon_{pq}^{(ij)}$ is the locally varying strain fields.\
The displacement fields $\chi^{kl}$ are found by solving the elasticity equation with a prescribed macroscopic strain, which in the discrete form reads:
\begin{equation}
    \mathbf{K}_s \bm{\chi}^m = \mathbf{F}^m, \hspace{1cm} \mathbf{F}^m = \sum_{e=1}^{N_e} \int_{\Omega_e} \mathbf{B}^T \mathbf{D} \bm{\epsilon}^m d \Omega_e,
\end{equation}
with $N_e$ the number of elements in the discretization, $\mathbf{K}_s$ the stiffness matrix, $\mathbf{B}$ the strain-displacement matrix, $\mathbf{D}$ the constitutive matrix and $\bm{\epsilon}^m$ the $m$th loading strain.\
The first versions of topology optimization, introduced by Bendsoe and Kikuchi~\cite{bendsoe1988generating}, were based on the homogenization method by representing each element in the design domain by equivalent homogenized properties.\
From early on in the field of density-based topology optimization, the homogenization theory is used to design periodic structures with certain desired constitutive properties~\cite{sigmund1995tailoring,hassani1998review,hassani2012homogenization}.\
This has lead today to a broad and still active field of research to find periodic media with exotic properties, e.g.~\cite{groen2018homogenization,zhang2019computational,osanov2016topology}.\

In this section, a focus is only put on the works who explicitly use the homogenization theory inside the topology optimization to obtain desired connectivity properties.\
Note that these methods only make sense to use when periodic (micro)structures are under investigation which assume infinite periodicity.\
Using equivalent thermal~\cite{sigmund1999optimality} or structural~\cite{sigmund1996composites} properties was already used in the early stages of topology optimization to control the connectivity of the solid or void phase.\ 
The work of Andreasen et al.~\cite{andreasen2014realization} is, however, generally seen as one of the first investigations where the connectivity issue is addressed explicitly in the text, and not just used without mentioning.\ 
Here, the authors used a homogenization-based constraint to achieve connectivity.\ 
Similar to the static compliance, the similarity between obtaining certain stiffness constitutive properties and the connectivity in the elastic phase was noticed.\
In their specific case of investigating two-phase viscoelastic composites, they however propose a connectivity constraint based on the minimum effective conductivity, to the example of Sigmund~\cite{sigmund1999optimality}.\
Equivalences exist between the bulk modulus and electrical/thermal conductivity for isotropic materials with real material properties~\cite{gibiansky1996connection,andreasen2014realization}.\
A lower bound on the conductivity can help assuring connectivity in the elastic phase by assigning high conductive material properties to the elastic phase while low properties to the other~phase.\
The authors, however, note that in 3D, isolated islands of materials can still occur.\

Over the years, the homogenization-based constraint was mostly proposed and applied during bandgap maximization (or corresponding wave attenuation maximization) of phononic or photonic crystals.\
Since the inherent nature of the achieved bandgaps rely on large differences in stiffness, the resulting topologies commonly consist of (near)-floating islands of solid material in the soft void phase.\
Using constraints on the equivalent stiffness properties can help ensuring a connected design is achieved.\
In a series of publications, Hedayatrasa et al.~\cite{hedayatrasa2016optimum, hedayatrasa2017maximizing, hedayatrasa2018optimization} presented  the bandgap maximization of phononic crystals while including the in-plane stiffness to obtain manufacturable designs as a second objective.\
Although presented in a GA optimization framework, they do not explicitly use the black-white nature of the underlying design variables.\ 
This in contrast to e.g.~\cite{bilal2012topologically,dong2014multi}, where the authors enforce connectivity with the explicit knowledge of the zero-white design field.\
Chen et al.~\cite{chen2019maximizing} presented an effective stiffness constraint based on the effective bulk modulus to control the load-bearing capabilities of phononic crystals while using the bi-directional evolutionary structure optimization.\
Although not the primary goal, the authors concluded that the constraint can help ensuring connectivity throughout the design.\
Rong et al.~\cite{rong2019topology} used an effective static modulus constraint in only one direction of periodicity to obtain elastic metamaterials which are fully connected, while Wang et al.~\cite{wang2022topological} applied a constraint on the effective stiffness moduli for the design of lattice metamaterials in an underwater environment.\
Swartz et al.~\cite{swartz2022manufacturing} recently investigate several connectivity constraint while maximizing the bandgap of photonic crystals.\
Both a constraint on the effective bulk modulus, effective shear modulus and a measure for anisotropy have been investigated.\
As in the earlier work of Andreasen et al.~\cite{andreasen2014realization}, they concluded that the effective bulk and shear moduli can help in obtaining certain desired properties regarding the stiffness and help removing isolated islands of material, but also that it does not guarantee that fully connected designs are achieved.\

\subsection{Geometric connectivity constraints}
\label{subsec:geom_constraint}
All connectivity constraints discussed in the previous section use an underlying physical phenomenon to obtain the desired connectivity.\
However, connectivity constraints have been proposed which are solely based on the underlying geometry/ topology without using equivalences based on physical phenomena such as conductivity or stiffness.\
This section discusses four different geometric connectivity constraints.\

\subsubsection{Void feature based}
A first step of geometric connectivity constraints can be seen as void feature based constraints.\ 
Instead of seeing the optimization as '\textit{where to place the material}', these methods reverse the logic of '\textit{where to place the void}' and control the placement of the void to restricted areas.\
Two categories can be distinguished: the side constraint method and projection-based constraint.\

\begin{figure*}
	\centering
	\includegraphics[width = \linewidth]{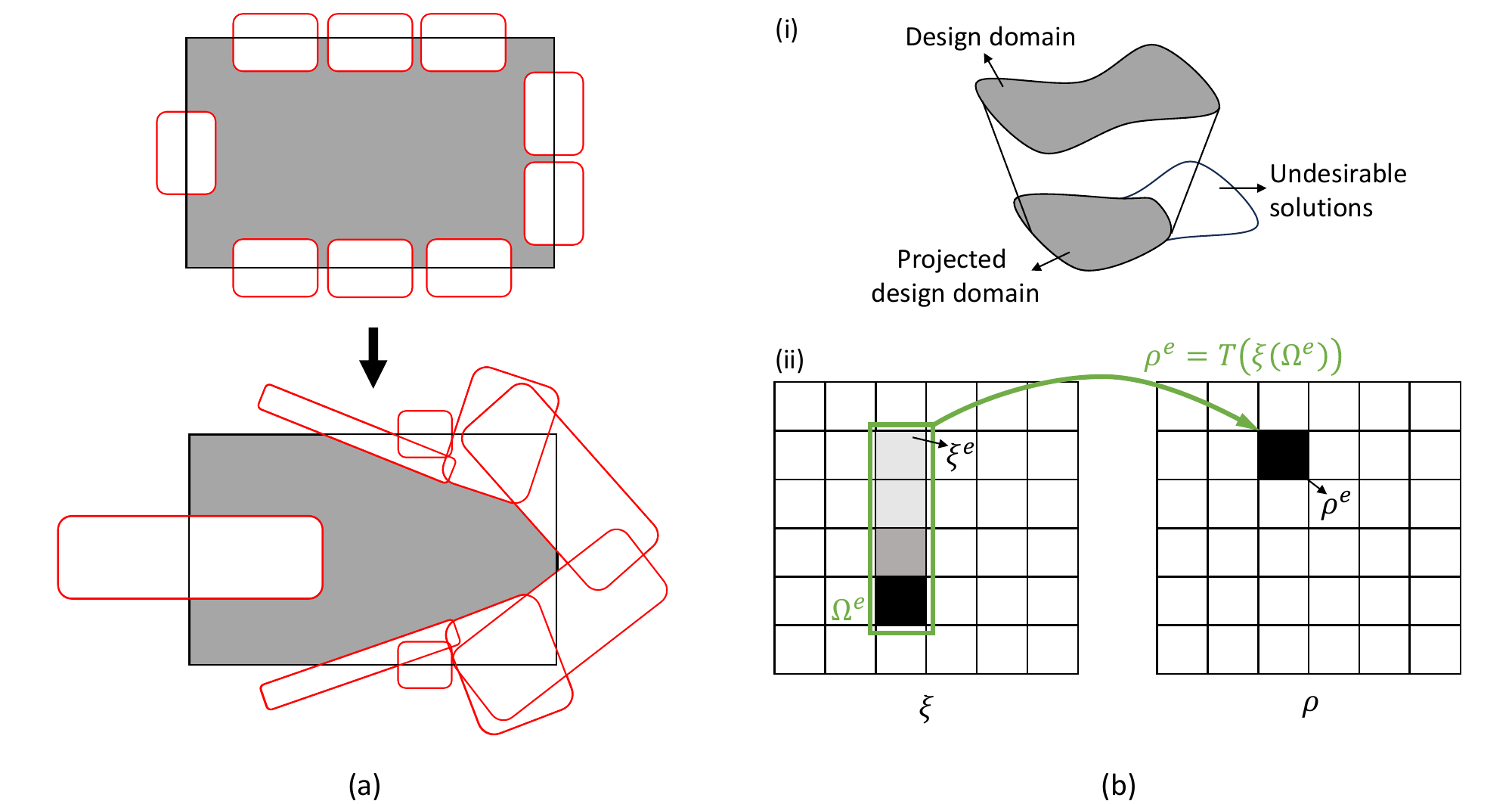}
	\caption{Schematic of the void feature based geometric connectivity constraints.\ a)~Visualization of the side constraint method in which the red features are controlled to determine the topology while constraining there center point on the boundary.\ Figure adapted from~\cite{zhou2019topology}.\ b)~Representation of the projection based method showing (i)~the limitation on the design domain and (ii)~an example of a projection with $T$ the mapping matrix and $\Omega^e$ the domain of influence for element $\xi^e$.\ Figure adapted from~\cite{vatanabe2016topology}. }
	\label{fig:VoidFeature}
\end{figure*}

\paragraph{Side constraint method}
The first void feature based geometric connectivity constraint includes the side constraint as proposed by Zhou and Zhang~\cite{zhou2019topology}.\
In a level-set based optimization framework, the topology is defined by changing the shape and size of predetermined void features.\
The void features are represented by B-splines and super-ellipses leading to a feature-driven method.\
Since there is explicit knowledge on the void features, constraints can be placed on the void centers to obtain only voids near the edges of the domain.\
This method is schematically visualized in Fig.~\ref{fig:VoidFeature}a.\
Note that this method could also readily be placed underneath the family of SCC methods as previously discussed but obviously requires feature parametrization of the design problem.\

\paragraph{Projection-based}
In a projection-based method, the set of design variables is converted with filters or projections to a set of pseudo-densities which inherently include the features of interest.\
The relation between the design variables and pseudo-densities is given by a variable mapping or projection function which embeds the information of the constraint.\
Due to the projection, the set of admissible designs decreases.\
The principle of the projection-based method is schematically shown in Fig.~\ref{fig:VoidFeature}b.\
Common for these methods is that they utilize and extend on the use of image processing based filters, e.g.~density filters~\cite{bourdin2001filters} and projection filters~\cite{guest2004achieving}.\
Projection-based constraints have been proposed mostly to include manufacturing specific knowledge, e.g.~ensuring castability~\cite{gersborg2011explicit}, avoiding overhang with an overhang projection~\cite{van2018continuous}, obtaining self-supporting structures~\cite{gaynor2016topology} or turning, casting, or forging constraints~\cite{vatanabe2016topology}.\

Gaynor et al.~\cite{gaynor2020eliminating} proposed a projection-based methodology to avoid enclosed voids, denoted as \textit{occluded voids} in their manuscript.\
Although focused on overhang and void islands, this work is typically categorized within the family of connectivity constraints since void islands of materials are avoided.\
The authors used the reasoning of changing the design problem such that the optimization determines where to place the voids and subsequently limiting where these voids may nucleate.\
This is, more specifically, done by combining the void projection method~\cite{sigmund2007morphology,guest2009topology} with a layer-by-layer overhang projection~\cite{gaynor2016topology} giving the following projection for element $e$:
\begin{equation}
    \rho_{void}^e = \rho_{void}^e(\phi(\psi)) = 1-H(\mu^e(\phi(\psi))),
\end{equation}
with $\rho_{void}^e$ the pseudo-void density corresponding to element $e$, $\mu^e$ the weighted average operator defined as a density filtering~\cite{bourdin2001filters}, $H$ the Heaviside projection~\cite{guest2004achieving} and $\phi(\psi)$ the overhang projection defined by:
\begin{equation}
    \phi^i=\psi^i H(\mu_s^i(\xi)) 
\end{equation}
for layer $i$ with $\xi$ the design variable field, $\psi^i$ indicating if material is desired at the location, $\mu_s^i$ a density filtering with uniform support region function weights and considering the area around the element as the support neighborhood set~\cite{gaynor2020eliminating}.\
Note that it is the combined void projection and overhang projection that leads to the elimination of the void islands.\

\subsubsection{Spectral graph theory}
\label{subsub:spectral_graph}
\begin{figure*}
	\centering
	\includegraphics[width = 0.95\linewidth]{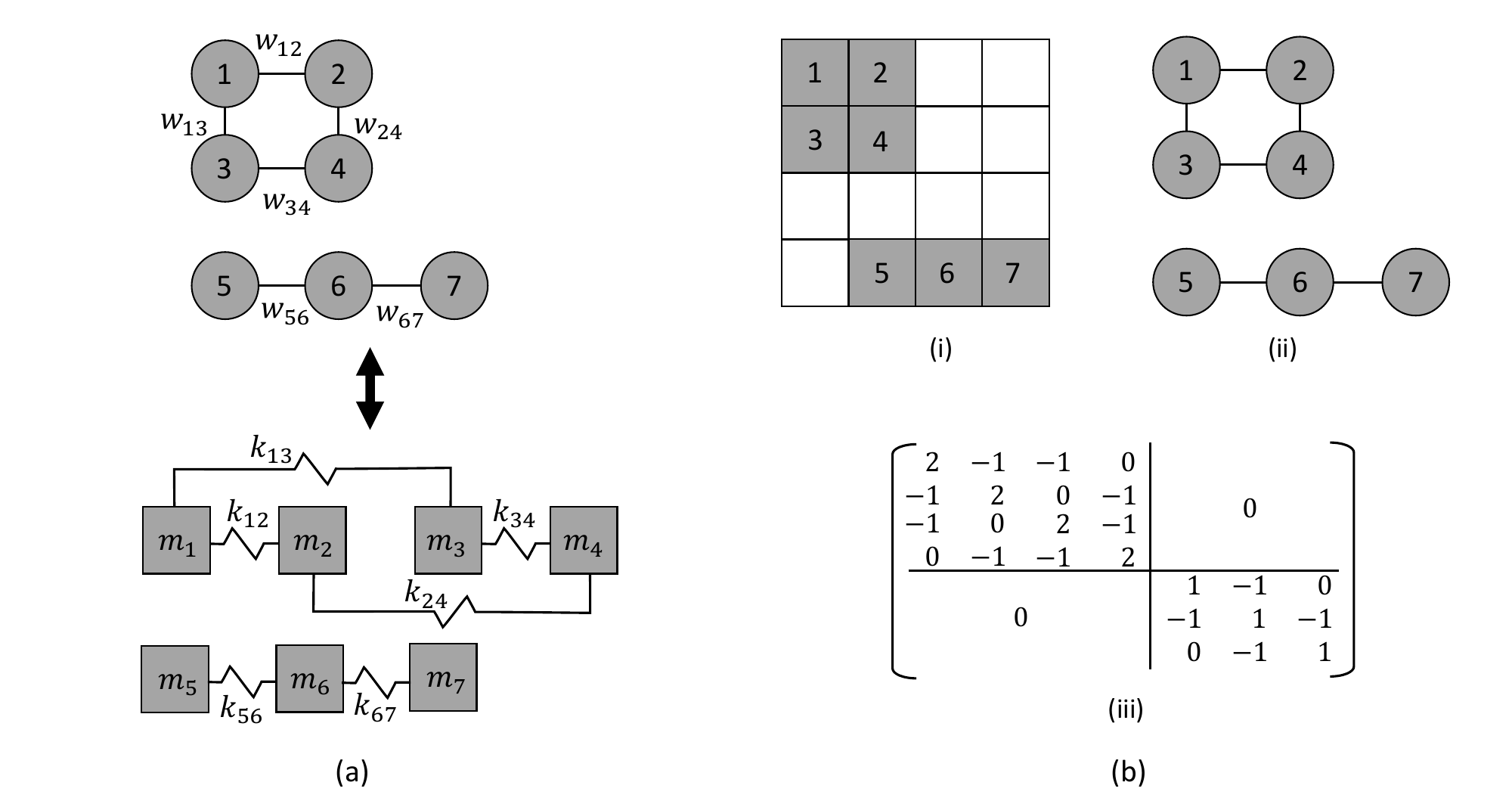}
	\caption{Visualization of the principle of the spectral graph theory-based connectivity constraint.\ a)~Equivalence between the weighted graph (top) and a discrete mechanical system of masses and springs (bottom).\ The nodes of the graphs can be seen as the masses, while the connections are the springs.\ b)~Example of the weighted graph and Laplacian matrix: (i)~example of a $4 \times 4$ design domain, (ii)~the corresponding graph representation and (iii)~the Laplacian matrix of the solid phase.\ Figures adapted from Donoso et al.~\cite{donoso2022new}. }
	\label{fig:Donoso_2022}
\end{figure*}

In recent years, Donoso et al.~have proposed a new set of connectivity constraints to avoid internal void holes based on spectral graph theory~\cite{chung1997spectral}.\
Firstly, a discrete form was proposed by representing the variable field as a weighted graph, i.e.~each element represents a node which is connected to the consecutive elements with a weighted positive number~\cite{donoso2022new}.\
The authors denote the equivalence of this representation with a discrete mechanical system, cf.~Fig.~\ref{fig:Donoso_2022}a.\
From the weighted graph, the Laplacian matrix $\bm{\Lambda}$ is constructed, cf.~the example of Fig.~\ref{fig:Donoso_2022}b, while the multiplicity of the zero eigenvalues of this matrix equals the number of connected components in the graph~\cite{fiedler1989laplacian}.\
Donoso et al.~propose to use the densities of the underlying variable field $\xi$ to define the weights $w_{ij}$ of the Laplacian matrix $\bm{\Lambda}$:
\begin{equation}
    w_{ij} = ((1-\xi_i)(1-\xi_j))^q (1-W_{min}) + W_{min},
\end{equation}
for $i,j$ adjacent elements, otherwise $w_{ij}$ equals 0.\ $q$ is a penalization term, $1-\xi$ is used since void islands of material were of interest by the authors and $W_{min}$ is a threshold to avoid null lines in the matrix.\
Next, the following eigenvalue problem is solved:
\begin{equation}
   (\bm{\Lambda}(\xi) -(\lambda_j -1)\mathbf{M}(\xi))\Phi_j=0,
\end{equation}
with $\mathbf{M}$ the system mass matrix, $\xi$ the design variable field and in which a shift of one on the eigenvalues is used.\
Using this construction, the number of void islands can be controlled by controlling the number of eigenvalues which equal 1 of the above problem.\
An equivalence can be seen with the eigenvalue-based connectivity constraints of Sec.~\ref{subsubsec:eig_constraint}, however the former mostly require the solution of a solid mechanics eigenvalue problem which is known to be computationally demanding while here a simpler scalar eigenvalue problem is solved.\ 
Note that the authors also propose to use an extra layer of void around the domain to ensure that a void space connected to the boundary is not detected as a void island.\
This method was applied for piezoelectric transducer design while ensuring connectivity in two phases simultaneously~\cite{donoso2023new}.\

Inspired by the discrete Laplacian matrix, the same authors recently proposed a rigorous continuous mathematical model to avoid void islands.\
The idea is to construct an auxiliary eigenproblem based on the 
Neumann-Laplacian problem~\cite{donoso2023continuous} or Dirichlet-Laplacian problem~\cite{donoso2024general}.\
From mathematics, it is known that the eigenvalues of these problems are related to the algebraic connectivity of the underlying design.\
The Dirichlet-Laplacian reads:
\begin{equation}
\label{eq:don_th}
\left\{
\begin{array}{ll}
-\mathrm{div}(w^x(\xi) \nabla \phi ) = \lambda m^x(\xi)\phi  \hspace{0.6cm} \mathrm{in} \hspace{0.1cm}\Omega \\ 
\phi=0 \hspace{4cm} \mathrm{on} \hspace{0.1cm} \partial\Omega 
\end{array}
\right. ,
\end{equation}
with $w^x$ and $m^x$ selected interpolation functions, here given if enclosed holes should be identified:
\begin{equation}
    \hspace{0.5cm} w^x(\xi) = (1-\epsilon)(1-\xi)^r+\epsilon,
    \hspace{0.5cm}  m^x(\xi) = 1-\xi,
\end{equation}
with $\epsilon$ a tuning parameter to avoid zero values and $r$ the penalty value.\
The corresponding connectivity constraint then reads: $\lambda > 0$.\
The use of the Dirichlet-Laplacian problem is preferred since it avoids the use of the additional void frame.\
The effectiveness of the constraint has been shown in both 2D and 3D finite component topology optimization~~\cite{donoso2023continuous,donoso2024general}.\
Although the constraints are mainly focused on avoiding disconnections in the void space, the principle is easily switched to avoiding solid islands of materials.\ 
This is applied as such in the practical comparison in this manuscript in the next section.\

\subsubsection{Cumulative sum flood fill}
\begin{figure*}
	\centering
	\includegraphics[width = \linewidth]{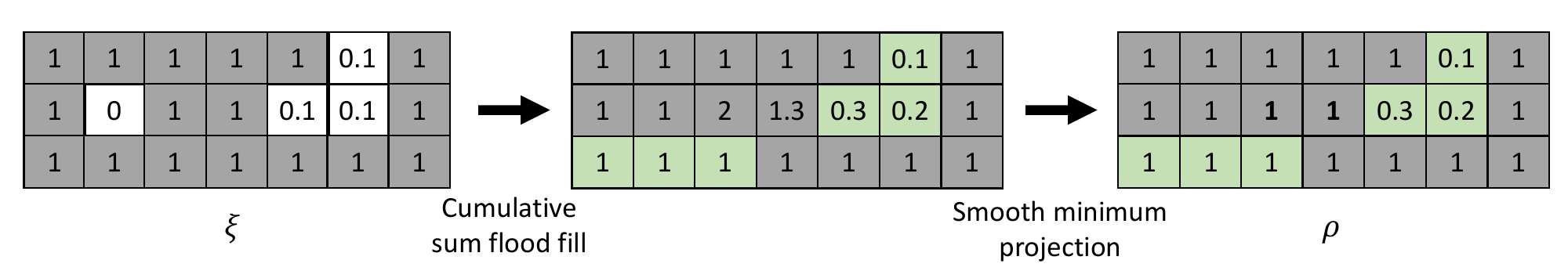}
	\caption{Example of the flood fill filtering as proposed by van der Zwet et al.~\cite{van2023prevention}. From left to right, first the initial design variable field is shown, next the design variable field after the filter operation and finally physical densities are retrieved with a smooth minimum projection method.\ Figure adapted from~\cite{van2023prevention}. }
	\label{fig:zwet}
\end{figure*}

Recently, van der Zwet et al.~\cite{van2023prevention} proposed a geometric-based connectivity constraint to remove enclosed voids in the design.\
The concept is based on the flood fill algorithm, commonly used in drawing programs~\cite{pavlidis1979filling}, combined with a cumulative sum approach, as presented by Langelaar~\cite{langelaar2019topology}.\
The concept of the flood fill algorithm is shown in Fig.~\ref{fig:zwet}.\
It takes the current design field as an input and loops over the different elements to detect and adapt the elements which are part of enclosed voids.\
Note that after the filtering, nonphysical densities above one are obtained, which are corrected with a smooth minimum projection.\
The connectivity constraint is then based on the difference between the current design field $\xi$ and the one obtained after the filtering:
\begin{equation}
    \sum_{e=1}^N \frac{(\rho_e-\xi_e)^2}{N} - \mu \leq 0
\end{equation}
in which $\rho_e$ is the value of element $e$ of the filtered design field and $\mu$ is a chosen threshold.\
The flood fill algorithm allows void passages with one single element, therefore the method should be combined by minimum length scale control in the void region e.g.~with morphology operators~\cite{sigmund2007morphology}.\
The authors also show that the amount of void channels can be controlled by using multiple independent flood fill steps on the initial unflooded design field.\
Since no additional PDE solution is required, this constraint comes with a lower computational cost compared to other physics-based constraints, but it should also be noted that efficient implementations require complex code.\
Moreover, in general one should not worry about solving a scalar PDE as this is usually orders of magnitude faster than the physical problem in question.\

\subsection{Overview}
\label{subsec:constraints_overview}
Now that the different connectivity constraints have been discussed, an overview is given in this section.\
Table~\ref{tab:overview} gives an overview on the capabilities of the discussed constraints.\
The gray coloring indicates for which sort of connectivity (cf.~Sec.~\ref{subsec:prop_conn}) the different methodologies have been presented and applied in the literature, while the usage of the symbol 'T' means the given method could be extended to handle the specific application.\ 
It is also indicated whether the constraint can be applied and/or is validated for 2D or 3D structures.\

A special note should be made regarding obtaining simply supported structures.\ 
Obtaining a simply supported structure generally involves the combination of avoiding void and solid islands of materials.\
If both void and solid islands can be detected, as indicated in the table, then theoretically also simply connected structures can be obtained.\
In specific cases, such as compliance minimization, avoiding void islands of material will result in simply connected structures since solid islands of material will not naturally appear.\
The reader should be aware of this when reading the literature since the usage of simply connected constraints is often used while only validating the methodologies with the simple compliance minimization.\

From the review in the previous sections and the overview table, it is clear that the physics-based constraints are mostly investigated.\ 
Only the VTM has been applied in all the definitions of connectivity, making it the most commonly used connectivity constraint.\
In the physics-based constraints, special attention should be placed on the selection of the boundary conditions since these will largely effect the influence the connectivity constraint has on the optimized design.\
This is particularly true for the static compliance constraint as this only ensures connection between loads and supports, and hence, does not prevent solid islands from appearing.\ 

From the physics-based constraints, the VTM has been firstly proposed to avoid void islands, while all others were originally proposed for the detection of solid islands of material.\
This is in contrast to the geometric constraints which are, until now, all proposed for the void island detection.\ 
Note here again that all the geometric constraint methodologies have only been applied for compliance minimization, except the discrete spectral graph method, making the structures simply-connected in the corresponding papers while no measures for the solid island removals were in place.\
The void feature and cumulative sum flood fill are also more difficult to extend towards the removal of solid islands since these methods rely on voids which accumulate from the boundaries of the domain.\ 
Using the same reasoning for the solid phase could still lead to loosely hanging solid structures at the boundaries of the design domain.\
It is also due to this reasoning that those methods will be harder to apply for periodic structures to avoid solid or void islands on the structure level.\

Lastly, a note is made on the eigenvalue-based and spectral graph methods.\
Due to the nature of these techniques, both methods are theoretically able to control the number of solid or void islands.\
This is possible by changing the eigenvalue which is controlled.\
Because of this, these methods could be categorized as well into the field of SCC methodologies, as discussed in Sec.~\ref{subsec:per_research}.\

Overall, it can be concluded from the table that possibilities for research are still present for the connectivity of periodic media, starting from the reference work of Swartz et al.~\cite{swartz2022manufacturing} and by extending the capabilities of the geometric constraints.\

\setlength{\arrayrulewidth}{0.5pt} 
\setlength{\cmidrulewidth}{0.5pt} 
\setlength{\aboverulesep}{0pt} 
\setlength{\belowrulesep}{0pt} 
\renewcommand{\arraystretch}{1} 
\begin{landscape}
\begin{table}[h]
\scalebox{0.9}{
  \centering
  \begin{tabular}{l | p{2cm}|p{2cm}|p{1.5cm} p{1.5cm}|p{1.5cm} p{1.5cm}| p{5.5cm}} 
   \toprule  
   & \multicolumn{2}{c|}{\textbf{Finite structures}} & \multicolumn{4}{c|}{\textbf{Periodic structures}} &  \\ 
   & Void island  & Solid island & \multicolumn{2}{c|}{Void island detection} & \multicolumn{2}{c|}{Solid island detection}  & Notes \\ 
   & \centering detection & \centering detection & \centering UC & \centering Structure & \centering UC & \centering Structure & \\
    \hline
    \multicolumn{8}{l}{\cellcolor{gray!40} \textbf{Physics-based constraints}} \\
    Poisson equation & &  & &  & &  & \\ 
    \hspace{0.5cm} VTM & \centering \cellcolor{gray!10} 2D, 3D & \centering \cellcolor{gray!10} 2D, 3D & \centering \cellcolor{gray!10} 2D, 3D & \centering \cellcolor{gray!10} 2D*, 3D* & \centering \cellcolor{gray!10} 2D, 3D & \centering \cellcolor{gray!10} 2D*, 3D* & * multiple UCs required\\ 
    \arrayrulecolor{gray!10} \cmidrule(r){2-8}
    \hspace{0.5cm} Other physics & \centering \cellcolor{gray!10} 2D, 3D &  \centering \cellcolor{gray!10} 2D, T3D & \centering T & \centering T* & \centering T & \centering T* & * multiple UCs required \\ 
    \arrayrulecolor{gray!40}\hline
    Eigenvalue based & &  & &  & &  & \\
    \hspace{0.5cm} Rigid body modes & \centering T$^\dagger$ & \centering \cellcolor{gray!10} 2D, 3D & \centering T & \centering T* & \centering \cellcolor{gray!10} 2D, 3D & \centering T* & * multiple UCs required \newline $\dagger$ T: Control on number of islands \\
    \arrayrulecolor{gray!20} \cmidrule(r){2-8}
    \hspace{0.5cm}  Gray penalization & \centering / & \centering \cellcolor{gray!10} 2D*, 3D* & \centering / & \centering / & \centering  T* & \centering / & * weakly connections detectable \\
     \arrayrulecolor{gray!40} \hline
    Equivalent performance & &  & &  & &  & \\
    \hspace{0.5cm} Static compliance$^{\dagger,\circ}$ & \centering T & \centering \cellcolor{gray!10} 2D, 3D & \centering T & \centering T* & \centering T & \centering T* & $\dagger$ mainly used for stiffness \newline $\circ$ only direct path between excitation and constraint guaranteed \newline * multiple UCs required \\
     \arrayrulecolor{gray!20} \cmidrule(r){2-8}
    \hspace{0.5cm} Self-weight compliance & \centering T & \centering \cellcolor{gray!10} 2D, T3D & \centering T & \centering T*  & \centering \cellcolor{gray!10} 2D$^\dagger$, 3D$^\dagger$ & \centering \cellcolor{gray!10} 2D$^\dagger$*,3D$^\dagger$* & * multiple UCs required \newline $\dagger$ only used for self-support \\
    \arrayrulecolor{gray!10} \cmidrule(r){2-8}
    \hspace{0.5cm} Homogenization$^{\dagger,\circ}$ & \centering / &  \centering / & \centering \cellcolor{gray!10} 2D, 3D & \centering \cellcolor{gray!10} 2D, 3D & \centering \cellcolor{gray!10} 2D, 3D & \centering \cellcolor{gray!10} 2D, 3D & $\dagger$ mainly used for stiffness \newline $\circ$ periodicity should be present \\
    \multicolumn{8}{|l|}{\cellcolor{gray!40} \textbf{Geometric constraints}} \\
   Void feature & &  & &  & &  & \\ 
   \hspace{0.5cm} Side constraint & \centering \cellcolor{gray!10} 2D, 3D & \centering  / & \centering T & \centering / & \centering / & \centering / & Compliance only tested \\ 
   \arrayrulecolor{gray!20} \cmidrule(r){2-8}
   \hspace{0.5cm} Projection-based & \centering \cellcolor{gray!10} 2D, 3D & \centering / & \centering T &  \centering / & \centering  / & \centering / & Compliance only tested \newline With overhang constraint   \\ 
   \hline
   Spectral graph & &  & &  & &  & \\ 
   \hspace{0.5cm} Discrete &\centering \cellcolor{gray!10} 2D*, 3D*  & \centering T* & \centering T &  \centering T & \centering  T & \centering T & * T: Control on number of islands \\ 
   \arrayrulecolor{gray!20} \cmidrule(r){2-8}
   \hspace{0.5cm} Continuous & \centering \cellcolor{gray!10} 2D*, 3D*  & \centering T* & \centering T &  \centering T & \centering  T & \centering T & Compliance only tested \newline * T: Control on number of island  \\ 
   \arrayrulecolor{gray!40}  \hline
  Cumulative sum flood fill & \centering \cellcolor{gray!10} 2D, 3D  & \centering / & \centering T &  \centering / & \centering  / & \centering / & Compliance only tested  \\ 
    \arrayrulecolor{black} \midrule
  \end{tabular}
  }
  \caption{Overview of the connectivity constraints for continuous design variables.\ The gray scaling indicates the connectivity constraint has been implemented and applied for 2D and/or 3D, while T represents the constraint could be used but that it has not been validated in the literature.\ For some constraints, special requirements hold which are indicated in the 'Notes' column.  }
 \label{tab:overview}
\end{table}
\end{landscape}

\section{Practical comparison}
\label{sec:comp}
The previous sections discussed the capabilities of the different connectivity constraints as present in the current state-of-the-art topology optimization literature.\
This section gives an elaborate comparison of five different connectivity constraints, both in terms of computational cost, the impact they have on the final obtained designs and on their respective performance characteristics.\
First the specific case study is discussed and optimized results are presented without considering the connectivity.\ 
Afterwards the five different investigated connectivity constraints are included and the findings are discussed.\ 

\subsection{Case study}
\begin{figure*}
	\centering
	\includegraphics[width = \linewidth]{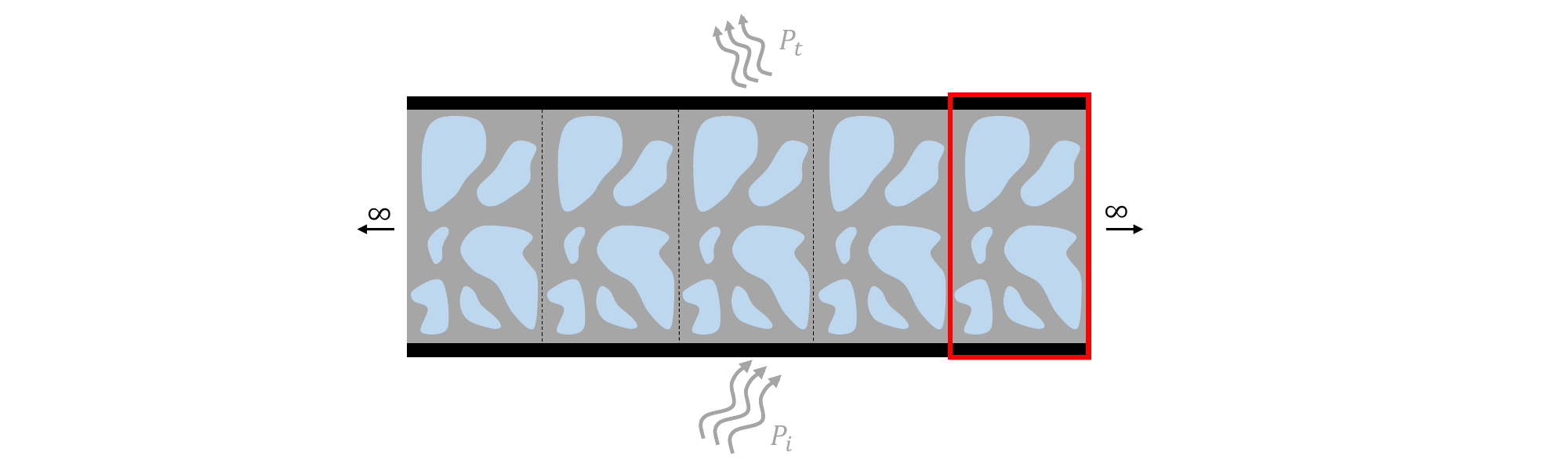}
	\caption{Overview of the problem under investigation during the comparison of the connectivity constraints.}
	\label{fig:scheme_stl}
\end{figure*}

As an application case, a vibroacoustic topology optimization framework for sandwich panel designs is selected~\cite{cool_TO_VA}.\
The problem under investigation is visualized in Fig.~\ref{fig:scheme_stl}.\
A double panel is considered with infinite periodicity in the horizontal direction while the core of one cell (shown in red) is the design domain.\
The panel is excited by a plane acoustic wave at the bottom with amplitude $P_i$ and the objective is to maximize the acoustic performance, i.e.~minimize the acoustic transmission, given in terms of the sound transmission loss (STL):
\begin{linenomath}
\begin{equation}
    \label{eq:STL}
    \mathrm{STL(\omega,\theta)} = -10\log_{10}{\left(\left| \frac{P_t}{P_i}\right|^2 \right)},
\end{equation}
\end{linenomath}
in which $\omega$, $\theta$ are, respectively, the frequency and angle of incidence, while $P_t$ is the amplitude of the transmitted pressure wave.\
The STL is computed using the wave and finite element (WFE) method~\cite{mace2008modelling}.\
This specific case is selected for two reasons: (i)~there is a need for a connectivity constraint between the top and bottom panel since it is known that a sandwich configuration with a fully acoustic core without structural connectivity between top and bottom has high STL~\cite{fahy2007sound}, (ii)~there is an interesting inherent trade-off between the desired acoustic performance and the required connectivity which influences the stiffness performance.\

A detailed presentation of the topology optimization, the sensitivity analysis, etc.~can be found in~\cite{cool_TO_VA}, whereas the following highlights the essentials in the context of connectivity constraints.\ 
The vibroacoustic coupling is included employing the method of Jensen~\cite{jensen2019simple}, while the robust formulation~\cite{wang2011projection} and double filtering technique~\cite{christiansen2015doublefilt} are used for regularization purposes~\cite{cool_TO_VA}.\
Due to the robust formulation, the optimization is written using a bound formulation as: 
\begin{subequations}
  \label{eq:TO_framework}
    \begin{empheq}{alignat=3} 
      & \min_{ \bm{\xi} \in \mathbb{R}^{N_e}, z \in \mathbb{R}^{+}} \quad & z  
        \label{eq:TO_a} \\
      & \hspace{0.3cm} \textrm{s.t.} &-\frac{\mathrm{STL}_e(\Delta\omega,\theta)}{C}+1 - z \leq 0 
        \label{eq:TO_b}\\
      & &-\frac{\mathrm{STL}_b(\Delta\omega,\theta)}{C}+1 - z \leq 0 
        \label{eq:TO_c}\\
      & &-\frac{\mathrm{STL}_d(\Delta\omega,\theta)}{C}+1 - z \leq 0
        \label{eq:TO_d}\\
      & &v_{d,\mathrm{P}}/V-1 \leq 0
        \label{eq:TO_e}\\
      & &z \geq 0, \hspace{0.3cm} 0 \leq \bm{\xi} \leq 1
        \label{eq:TO_f}
    \end{empheq}
\end{subequations}
in which $\bm{\xi}$ and $z$ are the $N_e+1$ design variables.\ 
Eqs.~(\ref{eq:TO_b}-\ref{eq:TO_d}) represent the acoustic performance with $C$ a scaling factor for normalization of the constraints.\ 
$\Delta \omega$ represents the frequency range of interest in which a number of discrete frequency lines are selected, $\theta$ is the angle of incidence and the subscripts $e,b,d$ denote the eroded, blueprint and dilated design of the robust formulation, respectively.\ 
Eq.~(\ref{eq:TO_e}) is a maximum volume constraint with $V$ the maximum allowed volume fraction.\ 
For completeness and convenience, additional details can be found in~\ref{app:TO_framework}.\ 

The periodic cell has dimensions $25\times 55$~mm with top/bottom plates of $2.5$~mm thickness.\ 
It is discretized with 2D bilinear plane strain elements of size $0.417$mm resulting in $7200$ design elements.\
A soft elastic Nylon-type plastic is used as material in the structural domain with Young's modulus $E=0.38(1+0.05\mathrm{i})$~GPa, density $\rho_s=1190$~kg/m$^3$ and Poissons ratio $\nu=0.35$.\ 
Air is assumed in the acoustic domain with density $\rho_a=1.225$~kg/m$^3$ and speed of sound $c_a=340(1+2\cdot10^{-4} \mathrm{i})$~m/s.\
The frequency range of interest is $2500-3000$~Hz, while the structure is excited by a normal plane wave and it should be noted that structural damping is introduced for both solid and fluid medium through complex material parameters.\
Further detailed characteristics of the problem and optimization settings are taken over from~\cite{cool_TO_VA}.\

\subsection{Result without connectivity constraint}
\begin{figure*}
	\centering
	\includegraphics[width = \linewidth]{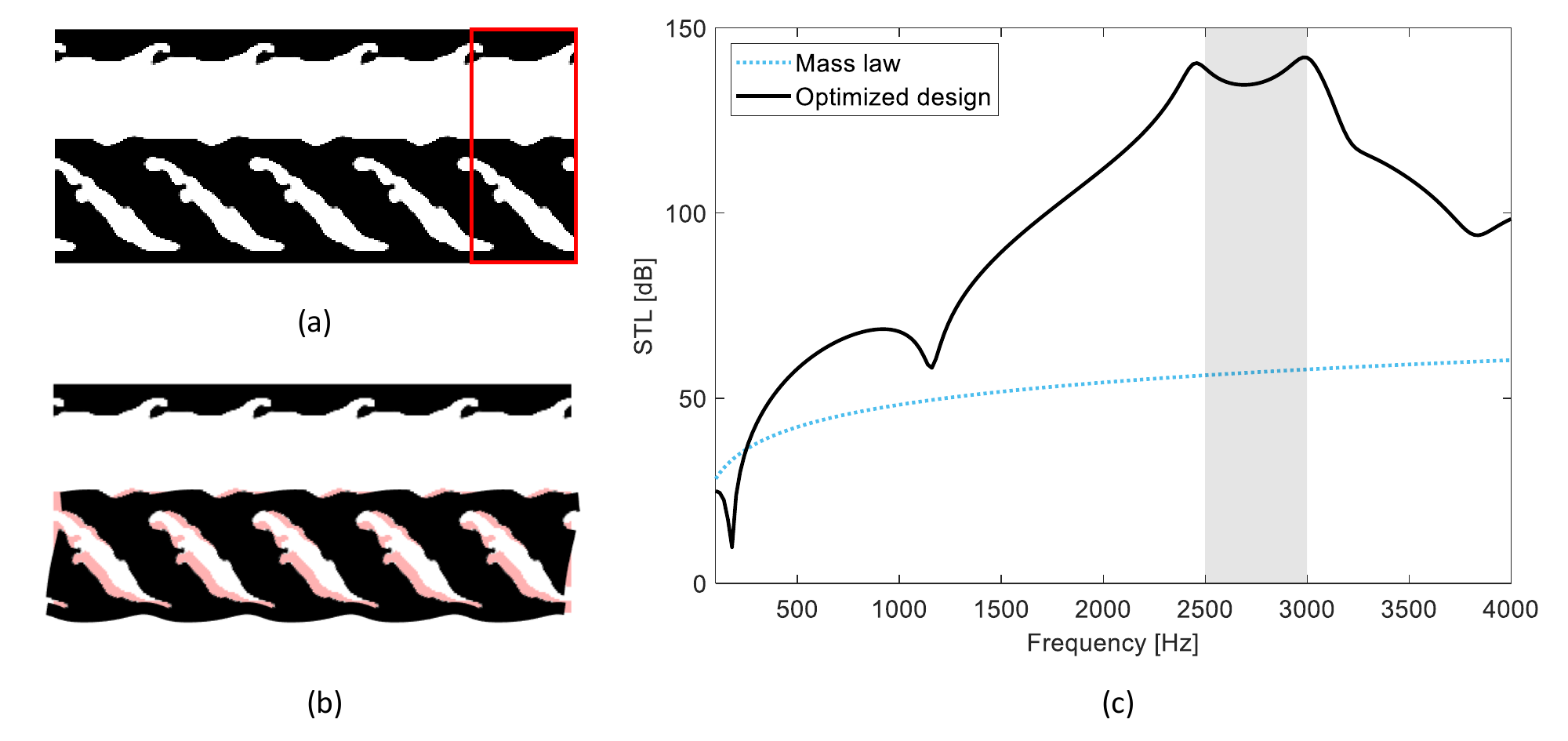}
	\caption{Result of the optimization problem without any connectivity constraint.\ a)~Visualization of the optimized sandwich panel consisting of five cells. The optimized cell is shown in the red square.\ b)~Vibration amplitude of the optimized design at $2750$~Hz.\ c)~STL performance of the optimized design together with the mass law as a reference case.\ The gray region indicates the considered frequency range during the optimization.  }
	\label{fig:uncoupled}
\end{figure*}

Solving the optimization problem of Eq.~(\ref{eq:TO_framework}), which does not include any constraint on the connectivity, results in the design of Fig.~\ref{fig:uncoupled}a where a clear separation of the top and bottom panel is visible.\
The final objective of this design equals $z=0.3245$.\
Fig.~\ref{fig:uncoupled}c shows a high STL performance is obtained in the optimized frequency range.\
Since the top and bottom panel are not structurally connected, the decoupling frequency of the sandwich panel occurs at a low frequency, i.e.~the STL dip at $180$~Hz.\ 
The displacement behavior in the optimized frequency range is visualized in Fig.~\ref{fig:uncoupled}b.\
This indeed shows that the bottom part of the structure has high displacements, while the top part has negligible movements resulting in the high STL.\
Since the top and bottom panels are not connected, this optimized structure is not manufacturable as one part.\ 
Moreover, the stiffness of the structure in the vertical direction is practically zero.\
This shows the need to enforce connectivity between the top and bottom panel.\

\subsection{Considered connectivity constraints}
Next a connectivity constraint\footnote{Note that the connectivity requirement can as well be added to the objective with a weight factor (instead of describing it as a constraint) to result in a multi-objective optimization.\ Changing the weight factor will give similar results and conclusions as changing the tolerance of the constraints.} is added to the optimization problem.\
As indicated before, the main goal of the connectivity constraint is to ensure that the top and bottom panels are connected such that a manufacturable design is achieved.\ 
Next, since the connectivity has a direct relation with the stiffness of the sandwich panel in the vertical direction, the connectivity constraint also serves as a way to control the load-bearing condition of the configuration.\

Five different constraints are compared in this section, chosen as those most commonly used in the literature.\
The connectivity constraints investigated are: 1)~the NVTM from Sec.~\ref{subsubsec:poisson_constraint}, 2)~the mechanical eigenvalue problem from Sec.~\ref{subsubsec:eig_constraint}, 3)~the static compliance constraint from Sec.~\ref{par:static}, 4)~the self-weight constraint from Sec.~\ref{par:self_weight} and finally 5)~the spectral graph theory-based constraint from Sec.~\ref{subsub:spectral_graph}.\

\begin{figure*}
	\centering
	\includegraphics[width = \linewidth]{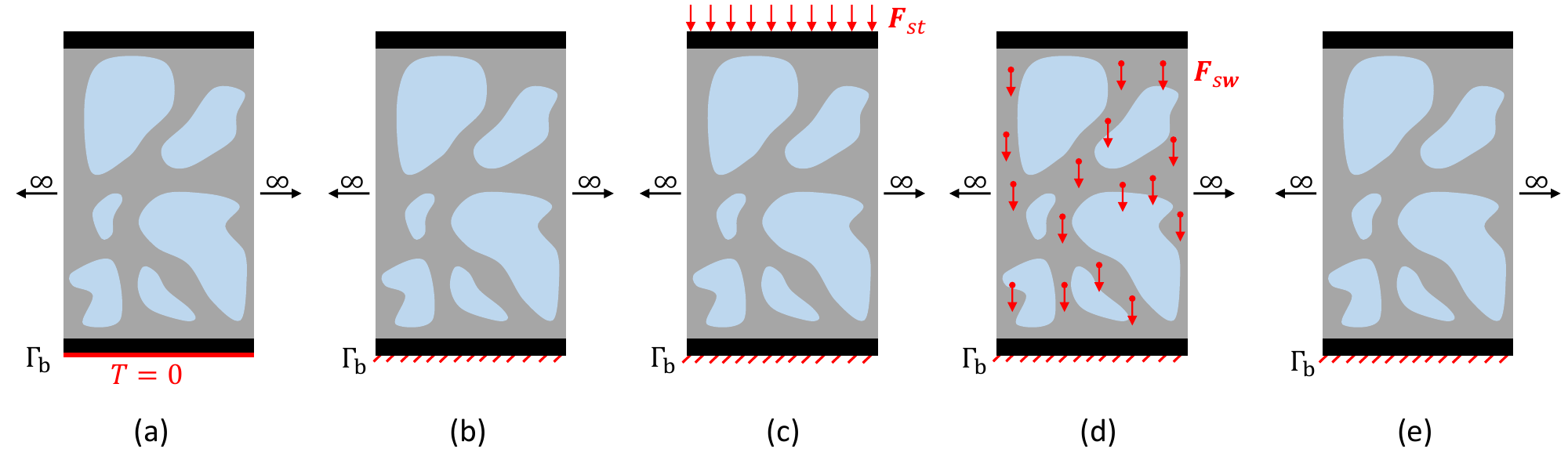}
	\caption{Overview of the boundary conditions for the different considered connectivity constraints in the comparison study: a)~NVTM, b)~mechanical eigenvalue, c)~static compliance, d)~self-weight compliance, e)~spectral graph connectivity constraint.}
	\label{fig:conn_cases}
\end{figure*}

General schematics of the different constraints along with the appropriate boundary conditions are given in Fig.~\ref{fig:conn_cases}.\
The NVTM (Fig.~\ref{fig:conn_cases}a) solves a thermal problem.\
The mechanical eigenvalue, static compliance and self-weight compliance  constraint (Fig.~\ref{fig:conn_cases}(b-d)) are defined using a purely structural FE since the assumption is made that the acoustic part of the design space will have a negligible contribution to the overall stiffness.\
The spectral graph constraint (Fig.~\ref{fig:conn_cases}e) solves an eigenvalue problem not representing any particular physics.\
Although different physical quantities are used to construct the constraints, the five constraints have some similarities in how they are defined for this specific case.\
All constraints are computed using one cell and using the physical density field representing the eroded design.\
To ensure the constraints are independent of the cell choice, i.e.~the connectivity constraint can be satisfied across the boundaries in the direction of periodicity, a periodicity boundary condition is applied in the horizontal direction.\
The density filtering used during the optimization routine is adapted accordingly to include this periodicity.\
Next, the different constraints are defined relative to a full design domain (design variables equal to one) to have a better scaling of the problem~\cite{swartz2022manufacturing}.\
Variables with an overhat notation mean they are computed on the full design domain.\
The particular constructions of the five different connectivity constraints for this problem are introduced with information necessary for the panel optimization problem at hand.\

\subsubsection{Nonlinear virtual temperature method}
The NVTM of Luo et al.~\cite{luo2020additive} is applied here to ensure a connected design between the top and bottom panel.\
One cell is illustrated, with $N_T$ DOFs, while periodic boundary conditions are applied in the $x$-direction, and a Dirichlet boundary condition $\mathbf{T}=\mathbf{0}$ is added only at the bottom (Fig.~\ref{fig:conn_cases}a):
\begin{equation}
\label{eq:NVTM}
\left\{
\begin{array}{ll}
    \mathbf{K}_T \mathbf{T} = \mathbf{Q}(\mathbf{T}) \hspace{2.2cm}    \mathbf{Q}(\mathbf{T}) = \frac{\mathbf{q}}{1+e^{\alpha(\mathbf{T}-\mathrm{T}_m)}} \\
    \mathbf{T}(x) = \mathbf{T}(x+nL_x) \hspace{1.35cm} n = \mathbb{Z} \\
    \mathbf{T}(x) = 0 \hspace{3cm} x \in \Gamma_b
\end{array}
\right.,
\end{equation}
in which $\mathbf{K}_T$ is the discretized system conductivity matrix.\
The sink boundary condition at the bottom is chosen since a heat source will always be present at the top (due to the passive top panel domain) and can only escape through the bottom (due to the $\mathbf{T}=\mathbf{0}$ Dirichlet boundary constraint).\ 
Hence, the constraint enforces a solid connection between top and bottom to limit the maximum temperature in the domain.\
As suggested in~\cite{luo2020additive}, the parameters are chosen as $\mathrm{T}_m = 100$ and $\alpha = 0.2$.\
The maximum temperature is determined using the $p$-norm with $p=12$ meaning that the connectivity can be ensured by constraining this maximum temperature relative to the temperature of a full solid design: 
\begin{equation}
        \theta_{vt} =  \left( \frac{1}{N_T}\sum^{n}_{i=1} \mathbf{T}_i^{p}\right)^{1/p}, 
        \hspace{1cm} \theta_{vt} \leq \mu_{vt} \hat{\theta}_{vt},
\end{equation}
with $\mu_{vt}$ a user-defined threshold value which will be varied in the numerical experiment and $\hat{\theta}_{vt}$ the maximum temperature observed when a full design domain is present.\
The sensitivities of this constraint are elaborated in~\ref{app:sens}.

\subsubsection{Mechanical eigenvalue}
The mechanical eigenvalue constraint is applied in this work not only to detect solid islands, but also to ensure a connected design between the top and bottom panel is obtained.\
One cell is envisaged while periodic boundary conditions are applied in the $x$-direction and the bottom is constrained (cf.~Fig.~\ref{fig:conn_cases}b):
\begin{equation}
\label{eq:eig}
\left\{
\begin{array}{ll}
(\mathbf{K}_s- \lambda_j \mathbf{M}_s)\phi_j = 0  \\
\phi_j(x) = \phi_j(x+nL_x) \hspace{1.25cm} n = \mathbb{Z} \\
\phi_j(x) = 0 \hspace{3cm} x \in \Gamma_b
\end{array}
\right. ,
\end{equation}
in which ($\lambda_j$, $\phi_j$) represent the $j^{th}$ eigenvalue and eigenvector.\
The eigenfrequency in Hz is defined as $\omega_j = \sqrt{\lambda_j}/2\pi$.\
Note that the zero boundary condition at the bottom is applied to increase the similarity to the other connectivity constraints.\
The ten lowest eigenvalues are computed and the minimum is obtained with a differentiable $p$-norm with $p = 12$ after which the connectivity constraint is defined relative to the full solid design:
\begin{equation}
        \theta_{eig} =  \left( \sum^{10}_{i=1} (\omega_i)^{-p}\right)^{-1/p}, \hspace{1cm}
         \theta_{eig} \geq \mu_{eig} \hat{\theta}_{eig},
\end{equation}
with $\mu_{eig}$ a user-defined threshold which will be varied in the numerical experiment and $\hat{\theta}_{eig}$ the minimum eigenvalue for a fully solid design.\
Since the top and bottom panels are passive domains and the bottom panel is constrained, a solid connection should be present to the top panel otherwise rigid body (zero eigenvalue) modes will appear violating the minimum allowed value as defined by $\mu_{eig} \hat{\theta}_{eig}$.\
The sensitivities of this constraint are elaborated in~\ref{app:sens}.\
Note that the $p$-norm is used instead of the first eigenvalue such that it is differentiable, even if degenerate eigenvalues are present~\cite{swartz2022manufacturing}.\

\subsubsection{Static compliance}
The connectivity between the top and bottom panels can be enforced by defining a particular static compliance problem.\
Since connectivity is desired in the $y$-direction, a uniform unit load is applied on the top panel in the negative $y$-direction ($\mathbf{F}_{st}$), while the bottom panel is constrained in both directions and periodicity boundary conditions are applied in the $x$-direction (cf.~Fig.~\ref{fig:conn_cases}c):
\begin{equation}
\label{eq:st}
\left\{
\begin{array}{ll}
\mathbf{K}_s \mathbf{U}_{st} = \mathbf{F}_{st} \\
\mathbf{U}_{st}(x) = \mathbf{U}_{st}(x+n L_x) \hspace{1.2cm} n = \mathbb{Z} \\
\mathbf{U}_{st}(x) = 0 \hspace{3.1cm} x \in \Gamma_b
\end{array}
\right. ,
\end{equation}
in which $\mathbf{U}_{st}$ is displacement vector.\
The static compliance, defined as $\theta_{st} = \mathbf{U}_{st}^T \mathbf{F}_{st}$, of this problem now embeds information on the connectivity between the top and bottom panel since the bottom panel is constrained.\
The static compliance of the envisaged problem will increase tremendously whenever the panels are not connected.\
Ensuring connectivity is now translated into obtaining a static compliance which does not exceed the static compliance of a full solid design domain ($\hat{\theta}_{st}$) multiplied by some threshold $\mu_{st}$:
\begin{equation}
        \theta_{st} \leq \mu_{st} \hat{\theta}_{st}.
\end{equation}
The threshold $\mu_{st}$ will be varied in the numerical experiment.\
The sensitivities of this constraint are given in~\ref{app:sens}.\
Note that this constraint enforces the top and bottom panel to be connected, but does not prevent the occurrence of additional islands of material.\

\subsubsection{Self-weight compliance}
Using a similar reasoning as in the previous section, the connectivity can as well be enforced using a self-weight compliance constraint.\
The same problem as for the static compliance is solved, only now a design-dependent body load is used.\
The self-weight force is applied in the negative $y$-direction while the bottom of the domain is constrained and periodicity boundary conditions are applied in the $x$-direction (cf.~Fig.~\ref{fig:conn_cases}d):
\begin{equation}
\label{eq:sw}
\left\{
\begin{array}{ll}
\mathbf{K}_s \mathbf{U}_{sw} = \mathbf{F}_{sw} = -\mathbf{N} \mathbf{x}_{Phys} \\
\mathbf{U}_{sw}(x) = \mathbf{U}_{sw}(x+n L_x) \hspace{1.2cm} n = \mathbb{Z} \\
\mathbf{U}_{sw}(x) = 0 \hspace{3.2cm} x \in \Gamma_b
\end{array}
\right. ,
\end{equation}
in which $\mathbf{N} \in \mathbb{R}^{N_s/2 \times N}$ contains zeros and minus ones to translate the element design variables towards structural nodal forces in the negative $y$-direction and $\mathbf{U}_{sw}$ is the DOFs vector with the $x$- and $y$-displacements.\
From this system, the self-weight compliance is defined and connectivity is enforced relative to the self-weight compliance of a full solid domain ($\hat{\theta}_{sw}$):
\begin{equation}
        \theta_{sw} = \mathbf{U}_{sw}^T \mathbf{F}_{sw}, \hspace{2cm} \theta_{sw} \leq \mu_{sw} \hat{\theta}_{sw},
\end{equation}
in which $\mu_{sw}$ is the threshold which will be varied in the numerical experiment.\
Whenever the structure is not self-supporting, $\theta_{sw}$ will become large.\
The sensitivities of this constraint are given in~\ref{app:sens}.

\subsubsection{Spectral graph connectivity constraint}
The spectral graph connectivity constraint of Donoso et al.~\cite{donoso2024general} has large similarities with the mechanical eigenvalue constraint.\ 
That is, one cell is considered while periodic boundary conditions are applied in the $x$-direction and the bottom is constrained (cf.~Fig.~\ref{fig:conn_cases}e), as done for the mechanical eigenvalue problem:
\begin{equation}
\label{eq:eig_geom}
\left\{
\begin{array}{ll}
(\mathbf{K}_d- (\lambda_j-\alpha_d) \mathbf{M}_d)\phi_j = 0  \\
\phi_j(x) = \phi_j(x+nL_x) \hspace{1.25cm} n = \mathbb{Z} \\
\phi_j(x) = 0 \hspace{3cm} x \in \Gamma_b
\end{array}
\right. ,
\end{equation}
in which ($\lambda_j$, $\phi_j$) represent the $j^{th}$ eigenvalue and eigenvector while $\alpha_d$ is a scalar equal to one~\cite{donoso2024general}.\
The system matrices $\mathbf{K}_d$ and $\mathbf{M}_d$ represent the discretized stiffness and mass system matrices of Eq.~(\ref{eq:don_th}), while the boundary conditions are in this case Dirichlet boundary conditions at the bottom and periodic boundary conditions at the left and right edge.\
The zero boundary condition at the bottom is applied to increase the similarity with the other connectivity constraints, while the periodic boundary conditions are added due to the infinite character of the structure in the horizontal direction.\
Note that the original Dirichlet-Laplace~\cite{donoso2024general} approach would suggest Dirichlet conditions on all edges but this would only avoid free floating features and not promote connectivity.\
For the interpolations $w^x$ and $m^x$, a RAMP interpolation gave the best results in this specific case:
\begin{equation}
    w^x =  m^x = \frac{\xi^e_\mathrm{P}}{1+q(1-\xi^e_\mathrm{P})} (1-\epsilon)+ \epsilon \hspace{1cm} 
\end{equation}
with $\xi^e_\mathrm{P}$ the physical design variables of the eroded design, $q=1$ and $\epsilon= 10^{-6}$.\
The five lowest eigenvalues are computed and the minimum is obtained using a $p$-norm with $p = 12$ after which the connectivity constraint is defined relative to the full solid design:
\begin{equation}
        \theta_{don} =  \left( \sum^{10}_{i=1} (\lambda_i)^{-p}\right)^{-1/p}, \hspace{1cm}
         \theta_{don} \geq \mu_{don} \hat{\theta}_{don},
\end{equation}
The sensitivities of the constraint are given in~\ref{app:sens}.

\subsection{Results}
The panel design problem is solved using the five different connectivity constraints defined in the previous section while the tolerances of the constraints are altered.\
This results in Pareto-graphs giving insight into the nature and impact of the individual connectivity constraints.\
The connectivity constraints are compared based on their consistency, Pareto-graphs and influence on the resulting designs and performances.\
However, first the computational cost of the constraints is examined for varying element numbers.\

\subsubsection{Computational cost}
The computational complexity of the different methods are as follows.\ 
The static and self-weight compliance require the solution of very similar problems, in which only the applied force changes.\
In both cases, a single system solution with sparse matrices and $2n_xn_y$~DOFs need to be solved, with $n_x$,$n_y$ indicating the number of nodes in the $x$- and $y$-direction, respectively.\
This system is solved with the backslash operator in Matlab.\
Since the problems are self-adjoint, no extra system solution is required for the sensitivities.\
For the self-weight compliance, a few extra matrix-vector or matrix-matrix operations are required to compute the design-dependent load and its sensitivity with respect to the design variables.\
For the mechanical eigenvalue constraint, a system with the same size ($2n_xn_y$~DOFs) is considered, however, now an eigenvalue problem needs to be solved.\
In this work, the eigenvalue problem is solved in Matlab with the iterative $eigs$-solver.\
Finally, the NVTM uses a thermal field resulting in a system of $n_xn_y$~DOFs.\
Since a non-linear system needs to be solved iteratively using Newtons method, multiple linear solves are required.\
Moreover, NVTM also requires the solution of a single additional adjoint problem.\
Again the Matlab backslash is used.\
Since a thermal problem is solved, the thermal system matrices need to be assembled specifically for this constraint.\
With a vector-implementation, this is, however, negligible.\
Finally, the spectral graph connectivity constraint also requires the solution of an eigenvalue problem.\ 
However, now only a scalar system of $n_xn_y$~DOFs should be solved in contrast to the mechanical eigenvalue constraint.\

\begin{figure*}
	\centering
	\includegraphics[width = 0.7\linewidth]{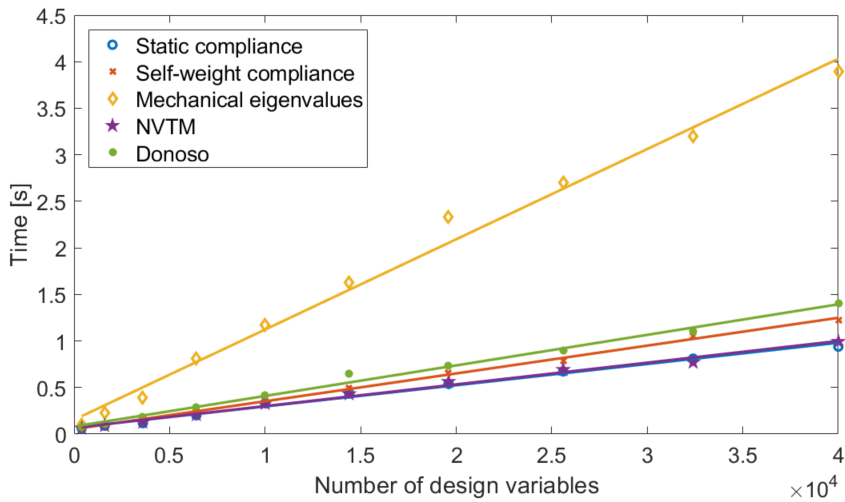}
	\caption{Analysis of computational cost for one constraint evaluation and corresponding sensitivity analysis for the five connectivity constraints.}
	\label{fig:comp_cost}
\end{figure*}

To quantify the computational costs in terms of wall-clock time, a single constraint and sensitivity evaluation is performed for varying number of elements~(Fig.~\ref{fig:comp_cost}).\
These calculations are executed on a laptop with 32GB RAM and a 2.6~GHz Intel Core i7-9540H processor.\
The different data points are plotted with a fitted curve through it.\
From the analyses, several conclusions can be drawn: (i)~due to the small number of design variables used in the numerical study, all constraints display a linear scaling, (ii)~as expected, the eigenvalue constraints (both mechanical and spectral graph) have the highest computational cost with the mechanical one largely above all others.\ The other three have a very similar computational cost.\ 
In conclusion, the mechanical eigenvalue constraint is not advisable in terms of computational cost, while the other constraints are comparable.\

\begin{figure*}[h!]
	\centering
	\includegraphics[width = \linewidth]{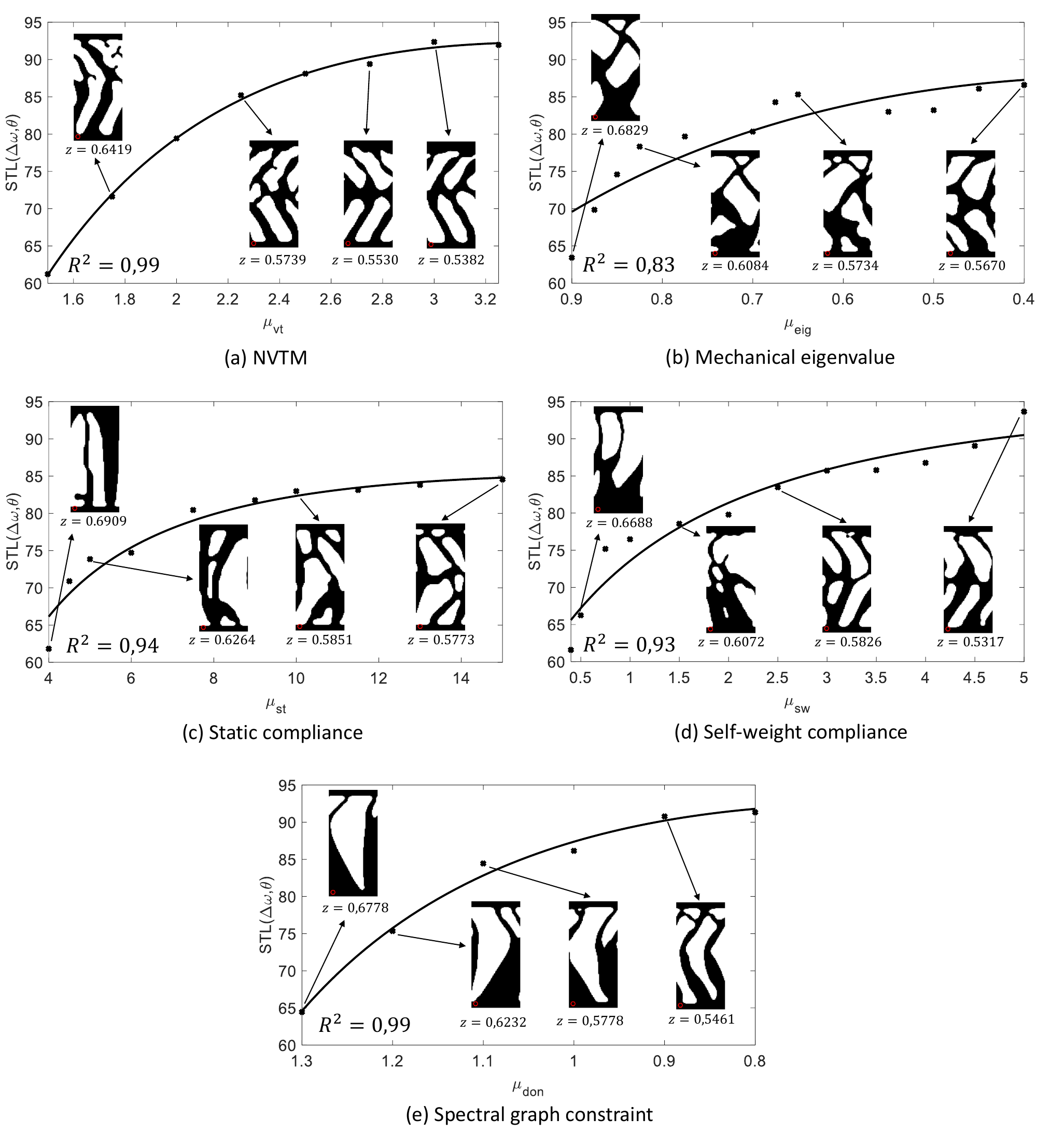}
 \caption{Results when changing the tolerance of the connectivity constraints.\ The Pareto graphs are visualized which show the STL performance versus the connectivity tolerance and four selected designs for: a)~NVTM, b)~mechanical eigenvalue, c)~static compliance, d)~self-weight compliance, e)~spectral graph connectivity constraint.\ The black solid line represents a fitted curve of the data points.\ The objective value $z$ for each design is given.\ Note that the result of the unconnected case is not added since it strongly outperforms the connected designs with $\mathrm{STL}(\Delta \omega, \theta) = 135.1$~dB.\ }
	\label{fig:pareto}
\end{figure*}

\subsubsection{Discussion Pareto-graphs}
Next, the Pareto-graphs of Fig.~\ref{fig:pareto} are discussed without focusing on the resulting designs.\
The Pareto-graphs are obtained as follows.\
For every $\mu$-value choice, the optimization algorithm is executed starting from six different initial guesses consisting of a random distributed gray core.\
The best result of these six is plotted as a dot in the graph.\
Afterwards, a fitting is done through the data points using the build-in $fit$-function of Matlab with the $Rat13$ model: a rational function of the form $(p_1 x+p_2)/(x^3+q_1 x^2+q_2 x+q_3)$.\
This fit is given as the black solid line in the graphs.\

From the resulting graphs in Fig.~\ref{fig:pareto}, different conclusions can be made.\
(i)~First, it is evident that the dependence of $\mu$ does not provide monotonic behavior of the objective function for all constraint types.\ 
Although this is to be expected due to non-convexity, monotonous behavior is desired and expected from an engineering perspective.\
Looking at the five different constraints, the scatter for the mechanical eigenvalue is the largest, followed by the compliance ones and finally the NVTM and spectral graph constraints.\
This is also reflected in the $R^2$-value of the fitted curves (Fig.~\ref{fig:pareto}).\
In summary, the NVTM has the most monotone behavior whereas the mechanical eigenvalue problem has the worst.\
(ii)~Secondly, and as expected, it is observed that a restrictive constraint leads to the worst STL performance.\ 
Note that this involves an increase in $\mu_{vt}$, $\mu_{st}$ and $\mu_{sw}$, but a decrease in $\mu_{eig}$ and $\mu_{don}$.\
More specifically, when moving from left to right, a steep increase is seen in the acoustic performance, after which this flattens out.\
Visually, one can say the flattening is more present for the static compliance and mechanical eigenvalue constraint, however this will be problem dependent.\
The Pareto-graphs show that a trade-off needs to be made between the desired STL performance and connectivity constraint in all cases.\
This will be further discussed in the next section.\
(iii)~Thirdly, the tunability of the constraints can be discussed.\
All envisaged $\mu$-ranges result in approximately the same optimized performance ($60-95$~dB), however, the values of the considered $\mu$-ranges differ significantly between the different constraints although all constraints are set-up relative to a reference solid design.\
The static compliance takes the biggest range, followed by the self-weight compliance, mechanical eigenvalue, spectral graph connectivity constraint and NVTM.\
This means that the resulting performance of the eigenvalue constraints and NVTM constraint is more sensitive towards a small change in $\mu$-parameter than the compliance ones.\

\subsection{Comparison tolerance values}
For each connectivity constraint, four $\mu$-values are selected covering the entire investigated $\mu$-range.\
For these selections, the corresponding $\mu$-value of the resulting optimized design is computed  for all other connectivity constraints.\
These values are summarized in Tab.~\ref{tab:tol}, in which the gray cells indicate the imposed $\mu$-values during the optimization and the other cells indicate the corresponding $\mu$-values for the other constraints.\
Also the STL performance ($STL(\Delta\omega,\theta)$ corresponding to the gray cell in each row is given for convenience.\
From this table, the different connectivity constraints are compared to each other based on their tolerance values and consistency between the constraints can be investigated.\

Per constraint (each time four rows), the values are given in terms of increasing acoustic performance, so decreasing objective.\
This results in an increase of $\mu_{vt}$, $\mu_{st}$ and $\mu_{sw}$, while a decrease of $\mu_{eig}$ and $\mu_{don}$ in the gray cells, as also indicated in the Pareto-graphs.\
Note that the $2.53$ value for $\mu_{vt}$ in the third row shows that in that specific case the connectivity constraint was not active in the end of the optimization since an original $\mu_{vt}$ of $3$ was imposed.\
Looking at the increasing or decreasing trend of the white cells in the table per four rows (the $\mu$-values computed in post-processing), observations can be made regarding the consistency between the different connectivity constraints.\
If in post-processing deviations are seen from the expected increase or decrease in $\mu$-values in the white cells of the table, it means there is no correspondence between that constraint and the one that was initially imposed (gray cells).\ 
From the table, the following observations can be made:
\vspace{-0.2cm}
\begin{itemize} [leftmargin=*]
    \item[-] The red rectangle in the table highlights the equivalence between the mechanical eigenvalue constraint, static compliance and self-weight compliance.\
    Using the mechanical eigenvalue constraint corresponds to imposing the static and self-weight compliance and the static compliance corresponds to imposing the self-weight compliance, i.e.~this is seen by the increasing fashion of $\mu$-values in the white cells above the gray ones in the red rectangle.\ 
    The other way around this observation does not hold, imposing the self-weight or static compliance does not translate into the equivalence with the mechanical eigenvalue, i.e.~the left bottom cells below the gray cells do not follow the expected increasing/decreasing pattern.\ 
    This can be understood from a physical point of view since the first targeted eigenvalue has a direct link to the stiffness of the structure, while a stiff structure in terms of the specific compliance can be obtained which results in a less interesting eigenmode.\ 
    \vspace{-0.2cm}
    \item[-] Looking at the NVTM constraint, it is seen that this does not explicitly correlate to the mechanical eigenvalue, static compliance nor self-weight compliance constraint in both directions (indicated by the blue rectangles in Tab.~\ref{tab:tol}).\ This is to be expected since the NVTM uses thermal analyses to impose the connectivity constraint while the other three use solid mechanic equivalences.\ It shows that the NVTM can be used to impose the connectivity, but will have a different physical influence on the results.\ 
    \vspace{-0.2cm}
    \item[-] For the spectral graph constraint, it is seen in the green rectangle in Tab.~\ref{tab:tol} that this one correlates to all the other constraints since the expected decrease in $\mu$ value is observed.\ Only in the self-weight compliance case a small deviation from this is seen.\
    This is an interesting observation showing the generality behind the methodology of the spectral graph connectivity constraint.\
    The other way around, however, does not hold, as seen in the last four rows of Tab.~\ref{tab:tol}.\
    For the NVTM, the influence seems to be even in the opposite direction since a decrease of $\mu_{vt}$ is obtained instead of an increase.\
\end{itemize}
\noindent
From above, it is clear that each discussed connectivity constraint is able to influence the strength of the connectivity.\ 
However, caution should be taken when choosing the connectivity constraint since inconsistencies are present between the constraints.\

\begin{table}[ht]
    \centering
    \begin{tikzpicture}
        \node at (0,0) {

         \begin{tabular}{c| c| c c c c c} 
          \arrayrulecolor{black}  \toprule  
             & STL & $\mu_{vt}$ & $\mu_{eig}$ & $\mu_{st}$  & $\mu_{sw}$  & $\mu_{don}$ \\
            \hline
            \multirow{4}{*}{NVTM} & 71.63 & \cellcolor{gray!40} 1.75 & 0.21 & 122.98 & 4.90 & 0.79 \\ 
             & 85.22 & \cellcolor{gray!40} 2.25 & 0.26 & 64.53 & 7.41  & 0.66\\ 
             & 89.40 & \cellcolor{gray!40} 2.75 & 0.18 & 193.49 & 16.55 & 0.55 \\ 
             & 92.36 & \cellcolor{gray!40} 2.53 & 0.19 & 91.01 & 15.16 & 0.51 \\ 
               \hline
            \multirow{4}{*}{Mechanical eigenvalue} & 63.42 & 3.82 & \cellcolor{gray!40} 0.9 & 14.32 & 0.84 & 0.97 \\ 
             & 78.33 & 3.88 & \cellcolor{gray!40} 0.825 & 18.34 & 1.23 & 0.87 \\
             & 85.32 & 3.81 & \cellcolor{gray!40} 0.65 & 19.84 & 2.21 & 0.74 \\
             & 86.60 & 3.40 & \cellcolor{gray!40} 0.4 & 34.19 & 4.22 & 0.68 \\
            \hline
            \multirow{4}{*}{Static compliance} & 61.82 & 2.63 & 0.17 & \cellcolor{gray!40} 4 & 1.13 & 0.63 \\ 
             & 73.85 & 4.07 & 0.27 & \cellcolor{gray!40} 6 & 1.54 & 0.63 \\
             & 82.99 & 3.60 & 0.40 & \cellcolor{gray!40} 10 & 2.05 & 0.62 \\
             & 84.55 & 3.44 & 0.37 & \cellcolor{gray!40} 14.97 & 3.60 &  0.54 \\
            \hline
            \multirow{4}{*}{Self-weight compliance} & 66.25 & 3.78 & 0.18 & 14.80 & \cellcolor{gray!40} 0.5 & 0.87 \\ 
             & 78.57 & 4.15 & 0.14 & 58.11 & \cellcolor{gray!40} 1.5 & 0.63 \\
             & 83.48 & 2.56 & 0.33 & 36.39 & \cellcolor{gray!40} 2.5 & 0.72 \\
             & 93.66 & 3.32 & 0.20 & 119.42 & \cellcolor{gray!40} 4.86 &  0.32 \\
            \hline
            \multirow{4}{*}{Spectral graph constraint} & 64.45 & 4.26 & 0.45 & 21.19 & 0.68 & \cellcolor{gray!40} 1.3 \\ 
             & 75.35 & 4.14 & 0.31 & 52.69 & 1.47 & \cellcolor{gray!40} 1.2 \\
             & 84.45 & 3.87 & 0.26 & 70.87 & 2.28 & \cellcolor{gray!40} 1.1 \\
             & 90.78 & 2.38 & 0.19 & 39.42 & 3.72 & \cellcolor{gray!40} 0.9 \\
             \midrule
          \end{tabular}
        };
        \draw[thick, line width=0.5mm, rounded corners=3pt, blue!30] (-0.2, -2.75) rectangle (0.7, 2.3);
        \draw[thick, line width=0.5mm, rounded corners=3pt, blue!30] (1, 2.4) rectangle (4.65, 4);
        \draw[thick, line width=0.5mm, rounded corners=3pt, red!40 ] (1, -2.75) rectangle (4.65, 2.3);
        \draw[thick, line width=0.5mm, rounded corners=3pt, green!30] (4.95, -2.75) rectangle (5.85, 4); 
    \end{tikzpicture}
    \caption{Comparison of the different connectivity constraints.\ For the four selected $\mu$-values for each connectivity constraint (gray cells), the equivalent $\mu$ is computed for all other considered connectivity constraints.\ The gray cells indicate therefore the imposed connectivity constraint and tolerance $\mu$, while the other $\mu$ values are computed in post-processing.\ For convenience, the corresponding acoustic performance $STL (\Delta\omega,\theta)$ is given in the table.\ The colored rectangles are solely used for indication purposes during the analysis of the table.\ }
     \label{tab:tol} 
\end{table}

\subsubsection{Influence on the resulting designs}
For each of the constraints, four designs are plotted in Fig.~\ref{fig:pareto} corresponding to four different $\mu$ values.\
From these results, the impact of the constraint on the optimized designs can be investigated.\

Looking at the designs where the connectivity constraint has most importance, i.e.~the left visualized design of each Pareto-graph in Fig.~\ref{fig:pareto}, the physical background of the different connectivity constraints becomes more clear.\
For the static compliance, a bulky design is obtained in which almost all material is allocated to form a rigid full solid connection between the panels.\
This is to be expected since as such the static compliance is minimized.\
For the self-weight constraint, it is seen that the structural material is allocated as much as possible close to the bottom panel, since the bottom panel is fixed, while only a slender connection to the top panel is present such that the top panel is supported as well.\
For the mechanical eigenvalue and spectral graph connectivity constraint, a similar behavior is seen, most of the material is allocated at the bottom, again due to the fixed bottom panel.\
For the mechanical eigenvalue, a cross-shape is recognized, which is a favorable structure to result in an as high as possible first mechanical eigenvalue while for the spectral graph connectivity constraint a bulky design at the bottom with slender connections to the top is achieved.\
Since the spectral graph connectivity constraint is a mathematical construction, the reason why these design phenomena occur is less intuitive than for the mechanical eigenvalue constraint.\
For the NVTM constraint, a design is obtained in which small features start to appear which are favorable for the heat dissipation.\

Whenever the connectivity constraint becomes of less importance, i.e.~moving to the right in the Pareto-graphs of Fig.~\ref{fig:pareto}, the above effects become less important.\
Although the same designs are not obtained between the five constraints, the optimized designs have similar features.\
This is to be expected noting that the constraint becomes less restrictive and the problem is prone to local minima.\
The core becomes more curved when moving to the right in the Pareto-graphs, while more slender features are obtained.\
Note that for each constraint except the spectral graph connectivity constraint, a design is present consisting of a circular core with curved connections to the top and bottom panel and a connection in the $x$-direction between consecutive UCs.\
Lastly, it is noted that the spectral graph connectivity constraint gives very similar designs for $\mu$ bigger than $1$ while large differences in the STL are obtained.\

A final observation related to the optimized designs regards the apparent length-scale perceived in the structures.\
Although the same minimum length-scale is enforced during the optimization independent of the selected connectivity constraint or $\mu$ value, it seems that some connectivity constraints lead to thicker features than others.\
This phenomenon is due to the underlying physical nature of the chosen connectivity constraint.\ The NVTM will make use of the minimum allowed length-scale since this is favorable for the conductivity of the structure.\ 
Using static compliance, slender features are inherently penalized in an intricate trade-off between the stiffness constraint and the needed acoustic flexibility.\
For the self-weight compliance, the minimum feature size is again noticed in the upper part of the structures since bulky parts get penalized due to their corresponding self-weight.\
For the mechanical eigenvalue and spectral graph connectivity constraint, the relationship is less straightforward due to the complex nature of the structures and corresponding eigenmodes.\

\begin{figure*}[h!]
	\centering
	\includegraphics[width = \linewidth]{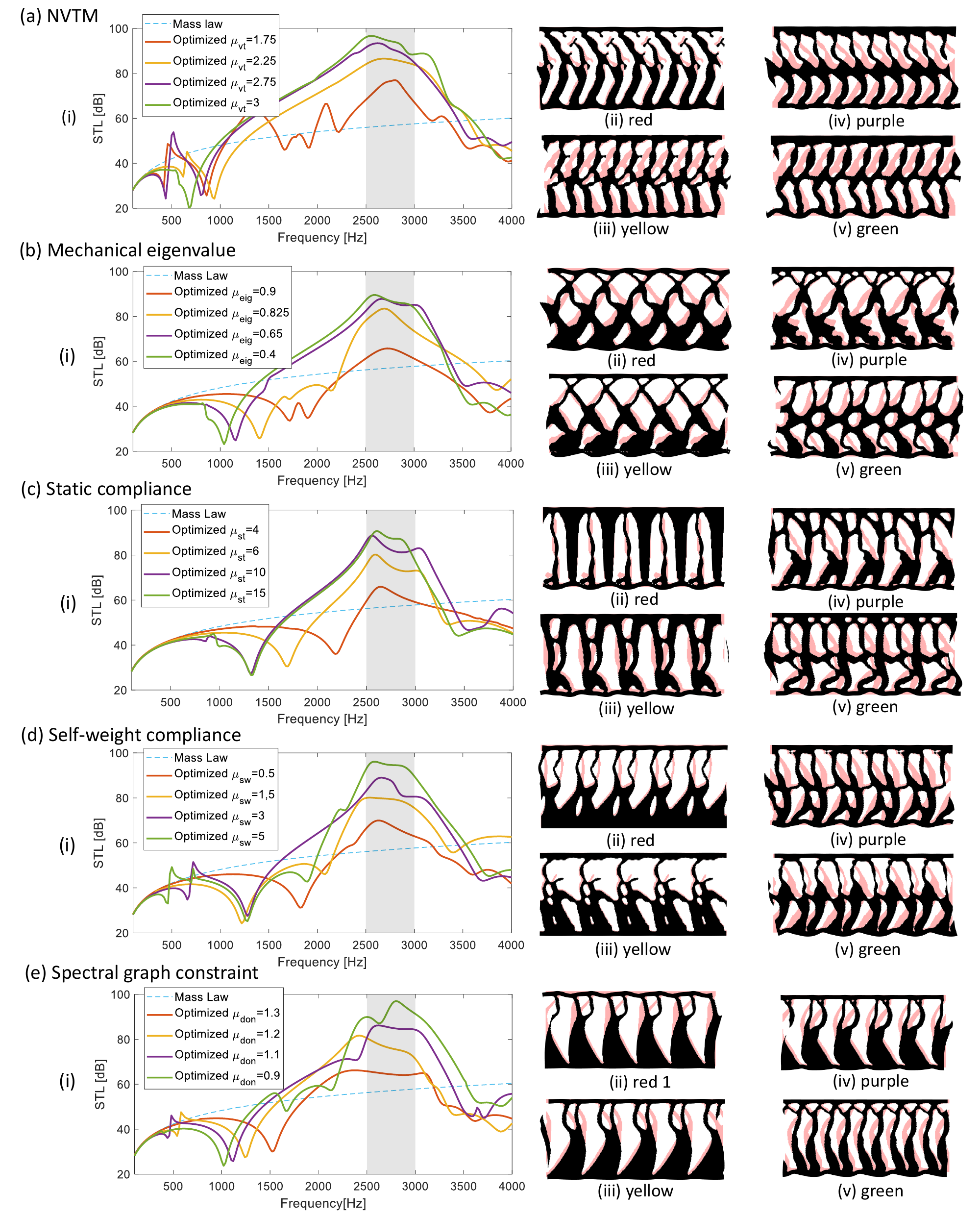}
	\caption{The corresponding STL (i) and vibration amplitude for five supercells visualized at $2750$~Hz (ii-v) for the designs visualized in Fig.~\ref{fig:pareto}: a)~NVTM, b)~mechanical eigenvalue, c)~static compliance, d)~self-weight compliance, c)~spectral graph connectivity constraint.\ Note that the result of the unconnected case is not added since it strongly outperforms the connected designs with $\mathrm{STL}(\Delta \omega, \theta) = 135.1$~dB.}
	\label{fig:stl}
\end{figure*}

\subsubsection{Influence on the STL performance}
The resulting STL performance of the highlighted designs in the previous section is computed over a broader frequency range, $100-4000$~Hz, and is plotted together with the vibration amplitude of the design at $2750$~Hz in Fig.~\ref{fig:stl}.\
This gives further insight into the exploited dynamics to obtain the STL when using the different constraints.\
It is noted that all obtained designs for this specific problem rely on mode-cancellation and mode-conversion, meaning the panels are decoupled in the targeted frequency range and the bending behavior of the bottom panel is translated to a translational motion of the top panel~\cite{cool_TO_VA}.\

First the case where the connectivity constraint is enforced the most, i.e.~red STL lines and corresponding designs in Fig.~\ref{fig:stl}(ii), are discussed.\ 
For the static and self-weight compliance constraint, it is seen that the STL performance has a dip-peak behavior.\
The decoupling frequency happens close to the envisaged frequency range, while the peak is obtained due to a local effect in the design, i.e.~a large displacement is seen in the slender connection in Fig.~\ref{fig:stl}a.ii and b.ii.\
For the mechanical eigenvalue and spectral graph connectivity constraints, a similar behavior is seen, while the peak is less pronounced and the vibration amplitude shows a more global displacement at the top side.\
In the NVTM, the STL curve takes a completely different shape.\
The design still relies on the decoupling happening before the frequency range of interest, however, many local peak responses are detected.\
This is a (favorable) consequence due to the small features which occur to obtain an as high as possible heat conduction which result in small resonators in the core.\

Moving to the case where the connectivity constraint becomes less important (purple and green), similar conclusions can be drawn for the static compliance, self-weight compliance and eigenvalue constraints: the decoupling frequency lowers with the connectivity constraint becoming less important while the STL has a broader response.\
The exploited dynamics are also similar, namely a combination of the mode-cancellation and mode-coupling effect in which a bending of the bottom panel, results in a rotation of the core and translation of the top panel \cite{cool_TO_VA}.\
For the NVTM constraint, a slightly different response is seen.\
The STL has also a broader characteristic whenever the constraint becomes less important.\
However, the shift in decoupling frequency as seen before is less pronounced in this case.\
This entails that the NVTM has less influence on the dynamic stiffness of the core in comparison to the other constraints.\
This complies with the expectations because it is the only constraint which uses another physical field to obtain the connectivity.\

\subsection{Overview}
The observations made in the previous sections are here summarized to get a good overview on how the different connectivity constraints compare to one another.\
Table~\ref{tab:discussion} gives an overview.\ 
The following conclusions can be drawn:
\vspace{-0.2cm}
\begin{itemize} [leftmargin=*]
    \item[-] Regarding the computational cost, the NVTM and static compliance constraint are most interesting to use while the one based on a mechanical eigenproblem has a high computational cost with respect to the other constraints.\
    \vspace{-0.2cm}
    \item[-] The variability, i.e.~monotonicity in objective with change in tolerance, of the connectivity constraint is most consistent for the NVTM and spectral graph constraint, while large scatter between different Pareto-points were observed when using the mechanical eigenvalue constraint.\
    \vspace{-0.2cm}
    \item[-] The static and self-weight compliance perform best in terms of tunability for the user since a larger $\mu$-range covers the Pareto-front, while small changes in $\mu$ for the NVTM and mechanical eigenvalue lead to large changes in achieved objectives.\
    \vspace{-0.2cm}
    \item[-] From a consistency analysis between the different constraints, the spectral graph connectivity constraint seemed to best correspond to the other constraints while the NVTM is least comparable to the other constraints.\
    \vspace{-0.2cm}
    \item[-] All constraints have a large impact on the resulting optimized designs.\ For the physics-based constraint, the underlying physics used to impose the connectivity is recognized in the optimized results and a strong dependency will be present on how the boundary conditions are selected.\ The spectral graph connectivity constraint seems to have a large impact on the resulting designs as well, since only smaller variations in the topology appeared when using the constraint compared to the other constraints.\
    \vspace{-0.2cm}
    \item[-] Regarding the impact on the performance, all constraints are comparable.\ Since the connectivity constraint was applied here with the second purpose of achieving structural integrity, the connectivity constraints using a physical reasoning closest to that purpose are preferable to choose.\ This, however, is dependent on the design problem at hand.\
\end{itemize}
\noindent
Overall, there is not one best connectivity constraint.\ 
Each constraint has its advantages and disadvantages.\ 
A careful selection should happen on which connectivity constraint to be used and the user should be aware on the large impact the constraint will have on the resulting optimized designs.\
Moreover, the boundary conditions applied in the connectivity constraints should be well thought off as well.\

\begin{table}[h]
  \centering
  \begin{tabular}{c| c c c c c } 
  \arrayrulecolor{black}  \toprule  
     &  \multirow{2}{*}{NVTM} &  Mechanical  &  Static   &  Self-weight  & Spectral graph  \\
     &  &  eigenvalue &  compliance  &  compliance & constraint \\
    \hline
    Computational cost &  ++ &  - &  ++ &  + & + \\ 
    Monotonicity &  ++ & - &  + &  + & ++ \\ 
    Tunability &  - &  - &  ++ & + & - \\ 
    Consistency  & - & + & - & - & ++ \\
    Impact design & -  & +/- & +/-  & +/- & - \\ 
    Impact performance &  +/- &  + &  + &  + & +/-  \\ 
     \midrule
  \end{tabular}
  \caption{ Comparison of the different investigated connectivity constraints in which the symbols $++$ until $-$ indicate a favorable til unfavorable behavior.}
 \label{tab:discussion}
\end{table}

\section{Conclusion}
\label{sec:concl}
This manuscript gives an overview and practical comparison of different available methodologies to control the \textit{connectivity} of the topology during a topology optimization process.\
Connectivity is defined as the control on the presence of enclosed void holes or solid islands of material.\
Focus is put on methods which are applicable when continuous design variables are used to represent the structure.\
The constraints are divided into two categories: physics-based and geometric connectivity constraints.\
The physics-based constraints are the most widely investigated type of constraints, with the virtual temperature method mostly applied.\
All physics-based constraints except the VTM have been introduced to avoid solid islands of material, while the geometric constraints have been presented focusing on the void islands.\
Until now, most geometric constraints have only been tested on compliance minimization problems, while their applicability for dynamic problems is yet to be demonstrated.\
Table~\ref{tab:overview} of this manuscript gives an overview on the applicability of the different connectivity constraints and the potential for extensions.\
Mostly for periodic media and the spectral graph geometric connectivity constraint there are interesting pathways to investigate in follow-up research.\

In a second part of the manuscript, a practical comparison of five different connectivity constraints is given.\ 
The NVTM, mechanical eigenvalue, static compliance, self-weight compliance and geometric spectral graph connectivity constraints have been applied to the optimization of a sandwich panel for its acoustic performance.\
From the analysis, it is concluded that the NVTM and static compliance are most interesting regarding the computational cost, while the spectral graph connectivity constraint is preferred in terms of the consistency with respect to the other constraints.\
The mechanical eigenvalue and self-weight compliance are interesting in terms of their implicit influence on the design and resulting performance due to their underlying used physical phenomena.\
An overview is given in Tab.~\ref{tab:discussion} of this manuscript.\
Care should be taken in the selection of the connectivity constraint since the physical background and choice of boundary conditions will largely impact the resulting optimized design.\

\section*{Acknowledgments}
The research of V.\ Cool (fellowship no.~1213925N) is funded by a grant from the Research Foundation - Flanders (FWO).\ 
The Research Fund KU Leuven is gratefully acknowledged for its support.\
O. Sigmund  and N. Aage acknowledge the financial support from the Villum Foundation through the Villum Investigator Project Amstrad (VIL54487).\ 
The resources and services used in this work were provided by the VSC (Flemish Supercomputer Center), funded by the FWO and the Flemish Government.\

\section*{Declarations}
\noindent
\textbf{Conflict of interest} On behalf of all authors, the corresponding author states that there is no conflict of interest.\\

\noindent
\textbf{Replication of results} The numerical examples in this study can all be reproduced using the procedures described in this paper. Furthermore, the data that support the findings of this study are available from the corresponding author upon reasonable request.

\newpage
\appendix

\section{Topology optimization framework}
\label{app:TO_framework}
This appendix elaborates on the optimization framework, while a complete description can be found in~\cite{cool_TO_VA}.

\subsection{From design variables to physical density fields}
The topology during the optimization is represented by the physical design variable field $\bm{\xi}_{\mathrm{P}}$ which is obtained from the design variables $\bm{\xi}$ using subsequent density filtering and Heaviside projections.\
A robust formulation is employed using an eroded~(e), blueprint~(b) and dilated~(d) design~\cite{wang2011projection}, together with the double filtering technique of~\cite{christiansen2015doublefilt}.\
The density filtering~\cite{bourdin2001filters} reads as follows:
\begin{linenomath}
\begin{equation}
\label{eq:dens}
        \Tilde{\xi}^e = \frac{\sum_{i=1}^{N_e} w(\xi^e - \xi^i)\xi^i}{\sum_{i=1}^{N_e} w(\xi^e - \xi^i)},
        \hspace{1cm}
         w(\xi^e - \xi^i) = \mathrm{max}\left(0, R-||\mathbf{x}^e-\mathbf{x}^i||_2 \right),
\end{equation}
\end{linenomath}
representing an averaging operation of the element design variables over a certain radius $R$ and $\mathbf{x}^j$ is the center position of element $j$.\
The Heaviside projection~\cite{guest2004achieving} pushes the design variables towards zero and one values and reads as follows: 
\begin{linenomath}
\begin{equation}
\label{eq:heav}
        \Bar{\xi}^e = \frac{\mathrm{tanh}(\eta\beta)+\mathrm{tanh}((\xi^e-\eta)\beta)}{\mathrm{tanh}(\eta\beta)+\mathrm{tanh}((1-\eta)\beta)},
\end{equation}
\end{linenomath}
with $\eta$ the projection level and $\beta$ the projection strength.\
The physical density fields $\bm{\xi}_{\mathrm{P}}$ are now obtained using the following operations:
\begin{linenomath}
\begin{equation}
\label{eq:double_filt}
    \bm{\xi} \hspace{0.2cm} \underset{R_1}{\overset{\mathrm{Eq}.~(\ref{eq:dens})}{\longrightarrow}} \hspace{0.3cm} 
    \Tilde{\bm{\xi}} \hspace{0.2cm} \underset{\eta_1,\beta_1}{\overset{\mathrm{Eq}.~(\ref{eq:heav})}{\longrightarrow}} \hspace{0.2cm} 
    \Bar{\Tilde{\bm{\xi}}} \hspace{0.2cm} \underset{R_2}{\overset{\mathrm{Eq}.~(\ref{eq:dens})}{\longrightarrow}} \hspace{0.2cm}
    \Tilde{\Bar{\Tilde{\bm{\xi}}}} \hspace{0.2cm} \underset{(\eta_e,\eta_b,\eta_d),\beta_2}{\overset{\mathrm{Eq}.~(\ref{eq:heav})}{\longrightarrow}} \hspace{0.2cm}
    (\Bar{\Tilde{\Bar{\Tilde{\bm{\xi}}}}}_e, \Bar{\Tilde{\Bar{\Tilde{\bm{\xi}}}}}_b, \Bar{\Tilde{\Bar{\Tilde{\bm{\xi}}}}}_d ) = (\bm{\xi}_{e,\mathrm{P}}, \bm{\xi}_{b,\mathrm{P}}, \bm{\xi}_{d,\mathrm{P}}),
\end{equation}
\end{linenomath}
with $(\eta_e,\eta_b,\eta_d) = (\eta_b+\Delta \eta,\eta_b,\eta_b-\Delta \eta)$.\

\subsection{Obtaining the system matrices}
With the physcial design fields, a material interpolation is performed per element to obtain the element material characteristics (Young's modulus $E^e$, density $\rho_s^e$, bulk modulus of the fluid $\kappa^e$ and density of the fluid $\rho_a^e$):
\begin{linenomath}
\begin{equation}
\label{eq:ADRAMP}
\left\{
\begin{array}{ll}
E^e(\xi^e_\mathrm{P}) = E_{v} + \frac{\xi^e_\mathrm{P}}{1+q(1-\xi^e_\mathrm{P})} (E-E_{v}),\\
\rho_s^e(\xi^e_\mathrm{P}) = \rho_{v} + \xi^e_\mathrm{P} (\rho_{s} -\rho_{v} ), \\
\frac{1}{\kappa^e(\xi^e_\mathrm{P})}  = \frac{1}{\kappa} + \xi^e_\mathrm{P} (\frac{1}{\kappa_r}-\frac{1}{\kappa}),  \\
\frac{1}{\rho_a^e(\xi^e_\mathrm{P})}  = \frac{1}{\rho_a} + \frac{\xi^e_P}{1+q(1-\xi^e_\mathrm{P})} (\frac{1}{\rho_r}-\frac{1}{\rho_a}),  \\
\end{array}
\right.
\end{equation}
\end{linenomath}
in which $E_v$, $\rho_v$, $\kappa_r$ and $\rho_r$ are artificial material properties to avoid numerical problems in the optimization.\
These interpolations represent a combination of the RAMP and linear interpolation~\cite{stolpe2001alternative}.\
Using the element material properties, the system mass, stiffness and coupling matrices are obtained by summing over all element contributions:
\begin{linenomath}
\begin{equation}
\label{eq:matrices}
\begin{aligned}
    &\mathbf{K}_s = \sum_{e=1}^{N_e} E^e(x,y) \mathbf{K}_s^e,
    \hspace{0.75cm}
    \mathbf{M}_s = \sum_{e=1}^{N_e} \rho_s^e(x,y) \mathbf{M}_s^e, \\
    &\mathbf{K}_a = \sum_{e=1}^{N_e} \frac{1}{\rho_a^e(x,y)} \mathbf{K}_a^e,
    \hspace{0.75cm} 
    \mathbf{M}_a = \sum_{e=1}^{N_e} \frac{1}{\kappa^e(x,y) } \mathbf{M}_a^e, \\
    &\mathbf{S}_p = -\sum_{e=1}^{N_e} (1-\xi^e_\mathrm{P}) \mathbf{S}^e,
    \hspace{1cm}
    \mathbf{S}_u = -\sum_{e=1}^{N_e} \xi^e_\mathrm{P} (\mathbf{S}^e)^T.
\end{aligned}
\end{equation}
\end{linenomath}
Note that the vibroacoustic coupling in the core is considered during the optimization using the method by Jensen~\cite{jensen2019simple}.\
The vibroacoustic coupled problem, representing the supercell, is given as follows:
\begin{linenomath}
\begin{equation}
\label{eq:VA_symm_exp}
    \left( -\omega^2 
    \underbrace{\left[ \begin{matrix} 
       \mathbf{M}_s       & 0\\
       \mathbf{S}_u         &   \mathbf{M}_a 
    \end{matrix} \right] }_{\mathbf{M}}
    +
    \underbrace{\left[ \begin{matrix} 
        \mathbf{K}_s       & \mathbf{S}_p\\
        0        &  \mathbf{K}_a 
    \end{matrix} \right] }_{\mathbf{K}} 
    \right)
    \underbrace{ \left[ \begin{matrix} 
        \mathbf{u} \\ \mathbf{p}
    \end{matrix} \right] }_{\mathbf{q}}
    =
    \underbrace{\left[ \begin{matrix} 
        \mathbf{f}_s \\ \mathbf{f}_a 
    \end{matrix} \right] }_{\mathbf{f}}
    +
    \underbrace{\left[ \begin{matrix} 
        \mathbf{e}_s \\ \mathbf{e}_a  
    \end{matrix} \right] } _{\mathbf{e}},
\end{equation}
\end{linenomath}
with the subscripts $s,a$ denoting the structural and acoustic domain, respectively.\
$\mathbf{M}, \mathbf{K}, \mathbf{S}$ represent the supercell system mass, stiffness and coupling matrices.\
Each node consists of three degrees-of-freedom (DOFs), being the $x$- and $y$-displacement, denoted as the vector $\mathbf{u}$ over all nodes, and the pressure, denoted as the vector $\mathbf{p}$ over all nodes.\
$\mathbf{q}, \mathbf{f}, \mathbf{e}$ represent, respectively, all DOFs of the supercell, the internal forces and external forces.\
Infinite periodicity is assumed in the horizontal direction, which is mathematically applied using the Bloch-Floquet boundary conditions~ \cite{bloch1929quantenmechanik}.\
Using the periodicity matrix $\mathbf{\Lambda}$, the nodal DOFs are related to the periodic DOFs vector $\hat{\mathbf{q}}$ which only contains the left and interior DOFs using the constant $\lambda_x = e^{-\mathrm{i} \mu_x}$ with $\mu_x = k_x L_x$ with $k_x$ , $L_x$ the wave propagation constant and length of the supercell, respectively, in the horizontal direction.\
This results in a modified system of equations~\cite{hussein2014dynamics}:
\begin{linenomath}
\begin{equation}
\label{eq:eq_motion_per_exp}
    (\hat{\mathbf{K}}-\omega^2 \hat{\mathbf{M}}) \hat{\mathbf{q}}=\hat{\mathbf{e}},
    \hspace{1cm}
    \hat{\mathbf{K}} = \mathbf{\Lambda}^H \mathbf{K} \mathbf{\Lambda},
    \hspace{1cm}
    \hat{\mathbf{M}} = \mathbf{\Lambda}^H \mathbf{M} \mathbf{\Lambda},
    \hspace{1cm}
    \hat{\mathbf{e}} = \mathbf{\Lambda}^H \mathbf{e},
\end{equation}
\end{linenomath}
Using this equation, the infinite STL corresponding to the supercell structure can be computed using the WFE technique~\cite{cool_TO_VA,boukadia2020wave}.\
For this, the surrounding acoustic halfspaces are modeled using an analytical formulation, while continuity conditions are applied to translate the structural displacement into the required pressure information.\

\subsection{Optimization problem}
The goal of the optimization is to maximize the acoustic performance in terms of the STL, resulting in a minimax problem due to the robust formulation:
\begin{linenomath}
\begin{equation}
\label{eq:minmax}
\min_{\bm{\xi}} \hspace{0.3cm} \max (-\mathrm{STL}_e(\Delta\omega,\theta),-\mathrm{STL}_b(\Delta\omega,\theta),-\mathrm{STL}_d(\Delta\omega,\theta)).
\end{equation}
\end{linenomath}
This minimax optimization problem is rewritten towards a bounded formulation while a constraint on the maximum allowed volume of the dilated design $v_{d,\mathrm{P}}$ is added: 
\begin{linenomath}
\begin{equation}
\label{eq:TO_framework_exp2}
\left\{
\begin{aligned}
\min_{ \bm{\xi} \in \mathbb{R}^{N_e}, z} \quad  z\\
\textrm{s.t.} \quad -\frac{\mathrm{STL}_e(\Delta\omega,\theta)}{C}+1 - z &\leq 0\\
 -\frac{\mathrm{STL}_b(\Delta\omega,\theta)}{C}+1 - z & \leq 0\\
 -\frac{\mathrm{STL}_d(\Delta\omega,\theta)}{C}+1 - z & \leq 0\\
 v_{d,\mathrm{P}}/V-1 & \leq 0\\
 z \geq 0, \hspace{0.3cm} 0 \leq \bm{\xi} &\leq 1,\\
\end{aligned}
\right.
\end{equation}
\end{linenomath}
which is the optimization problem presented in Eq.~(\ref{eq:TO_framework}).\

\section{Sensitivities}
\label{app:sens}
This section gives an overview of the sensitivities of the five different connectivity constraints which are used in the practical comparison.\
\subsection{Nonlinear virtual temperature method}
The sensitivity of $\theta_{vt}$ towards the design variables can be obtained using the adjoint method:
\begin{equation}
        \frac{d\theta_{vt}}{d \xi_j} = \Lambda_{vt}^T\left(\frac{d\mathbf{K}_T}{d \xi_j}\mathbf{T}-\frac{d\mathbf{Q}}{d \xi_j}\right),
\end{equation}
with $\Lambda_{vt}$ the adjoint vector and:
\begin{equation}
\label{eq:adjoint_nvtm}
    \left(\mathbf{K}_T-\frac{d\mathbf{Q}}{d\mathrm{T}}\right)\Lambda_{vt} = \left(-\frac{d\theta_{vt}}{d\mathrm{T}}\right)^T,
    \hspace{1cm}
    \frac{d\theta_{vt}}{d\mathrm{T}} = \theta_{vt}^{1-p}\frac{\mathbf{T}^{p-1}}{N_T}.
\end{equation}
The reader is referred to~\cite{luo2020additive} for more details.

\subsection{Mechanical eigenvalue}
The gradient of $\theta_{eig}$ towards the design variables can be obtained using the chain rule:
\begin{equation}
\begin{split}
    \frac{d\theta_{eig}}{d \xi_j} & = \frac{d\theta_{eig}}{d\omega_j} \frac{d\omega_{j}}{d\lambda_j} \frac{d\lambda_j}{d \xi_j},
    \\
     \frac{d\theta_{eig}}{d\omega_j} = \left(\sum_k \omega_k^{-p} \right)^{-1/p-1} & \omega_j^{-p-1}, 
     \hspace{1cm} 
     \frac{d\omega_{j}}{d\lambda_j} = \frac{1}{4\pi} (\lambda_j)^{-1/2},
     \\
     \frac{d\lambda_j}{d \xi_j} = & \phi_j^* \left( \frac{d\mathbf{K}_s}{d \xi_j}-\lambda_j \frac{d\mathbf{M}_s}{d \xi_j} \right) \phi_j,
\end{split}
\end{equation}
in which the asterisk indicates the Hermitian transpose.\ 
Finally the normalized connectivity constraint and its sensitivity read:
\begin{equation}
    g_{don} = \frac{\mu_{eig} \hat{\theta}_{eig}}{\theta_{eig}} \hspace{2cm} \frac{dg}{d \xi_j} = -\frac{\mu_{eig} \hat{\theta}_{eig}}{\theta_{eig}^2} \frac{d\theta_{eig}}{d \xi_j}
\end{equation}

\subsection{Static compliance}
The derivative of the static compliance $\theta_{st}$ is well known in the literature and efficient to compute due to the self-adjoint nature of the problem:
\begin{equation}
    \frac{d\theta_{st}}{d \xi_j} = -\mathbf{U}_{st}^T \frac{d\mathbf{K}_s}{d \xi_j} \mathbf{U}_{st},
\end{equation}
in which $d\mathbf{K}_s/d \xi_j$ is obtained as in standard topology optimization routines~\cite{bendsoe1999material}.\

\subsection{Self-weight compliance}
The gradient of $\theta_{sw}$ can be obtained efficiently due to the self-adjoint nature of the problem:
\begin{equation}
    \frac{d\theta_{sw}}{d \xi_j} =2\mathbf{U}^T_{sw} \left( \frac{d\mathbf{F}_{sw}}{d \xi_j} - \frac{d\mathbf{K}_s}{d \xi_j} U \right)
    \hspace{1cm}
    \frac{d\mathbf{F}_{sw}}{d \xi_j} = -\mathbf{N}_j,
\end{equation}
in which $\mathbf{N}_j$ represents the $j^{th}$ column of the matrix $\mathbf{N}$.

\subsection{Donoso connectivity constraint}
The gradient of $\theta_{don}$ towards the design variables can be obtained using the chain rule similar as for the mechanical eigenvalue:
\begin{equation}
\begin{split}
    \frac{d\theta_{don}}{d \xi_j} & = \frac{d\theta_{don}}{d\lambda_j} \frac{d\lambda_j}{d \xi_j},
    \\
     \frac{d\theta_{don}}{d\lambda_j} = \left(\sum_k \lambda_k^{-p} \right)^{-1/p-1}  \lambda_j^{-p-1}, &
     \hspace{1cm} 
     \frac{d\lambda_j}{d \xi_j} =  \phi_j^* \left( \frac{d\mathbf{K}_d}{d \xi_j}-(\lambda_j -\alpha_d) \frac{d\mathbf{M}_d}{d \xi_j} \right) \phi_j,
\end{split}
\end{equation}
in which the asterisk indicates the Hermitian transpose.\ 
Finally the normalized connectivity constraint and its sensitivity read:
\begin{equation}
    g_{don} = \frac{\mu_{don} \hat{\theta}_{don}}{\theta_{don}} \hspace{2cm} \frac{dg_{don}}{d \xi_j} = -\frac{\mu_{don} \hat{\theta}_{don}}{\theta_{don}^2} \frac{d\theta_{don}}{d \xi_j}
\end{equation}

\bibliography{mybibfile}

\end{document}